\documentclass[12pt,preprint]{aastex}

\shorttitle{Cool White Dwarfs in SDSS}
\shortauthors{Kilic et al.}
\begin{document}
\title{Cool White Dwarfs in the Sloan Digital Sky Survey\altaffilmark{1,2}}

\author{Mukremin Kilic\altaffilmark{3,7}, Jeffrey A. Munn\altaffilmark{4}, Hugh C. Harris\altaffilmark{4}, James Liebert\altaffilmark{5}, Ted von Hippel\altaffilmark{3}, Kurtis A. Williams\altaffilmark{5}, Travis S. Metcalfe\altaffilmark{6}, D. E. Winget\altaffilmark{3}, and Stephen E. Levine\altaffilmark{4}}

\altaffiltext{1}{Based on observations obtained with the Hobby-Eberly Telescope, which is a joint project of the University of Texas at Austin, the Pennsylvania State University, Stanford University, Ludwig-Maximilians-Universit\"at M\"unchen, and Georg-August-Universit\"at G\"ottingen.}
\altaffiltext{2}{Observations reported here were obtained at the MMT Observatory, a joint facility of the Smithsonian Institution and the University of Arizona.}
\altaffiltext{3}{The University of Texas at Austin, Department of Astronomy, 1 University Station C1400, Austin, TX 78712}
\altaffiltext{4}{US Naval Observatory, P.O. Box 1149, Flagstaff, AZ 86002}
\altaffiltext{5}{Steward Observatory, University of Arizona, 933 North Cherry Avenue, Tucson, AZ 85721}
\altaffiltext{6}{Harvard-Smithsonian Center for Astrophysics, Mail Stop 16, 60 Garden Street, Cambridge, MA 02138}
\altaffiltext{7}{kilic@astro.as.utexas.edu}

\begin{abstract}
A reduced proper motion diagram utilizing Sloan Digital Sky Survey (SDSS)
photometry and astrometry and USNO-B plate astrometry is used
to separate cool white dwarf candidates from metal-weak, high-velocity main
sequence Population II stars (subdwarfs) in the SDSS Data Release 2 imaging area.
Follow-up spectroscopy using the Hobby-Eberly Telescope, the MMT, and
the McDonald 2.7m Telescope is used to demonstrate that the white dwarf and
subdwarf loci separate cleanly in the reduced proper motion diagram, and
that the contamination by subdwarfs is small near the cool white dwarf locus.
This enables large statistically complete samples of white dwarfs, particularly
the poorly understood cool white dwarfs, to be created from the SDSS imaging
survey, with important implications for white dwarf luminosity function
studies.  SDSS photometry for our sample of cool white dwarfs is compared to
current white dwarf models.
\end{abstract}

\keywords{stars: atmospheres---stars: evolution---white dwarfs}

\section{Introduction}
The white dwarf luminosity function of the Galactic disk has traditionally been used
as one tool to estimate the star formation history and age of this
population (Liebert 1979; Winget et al. 1987; Liebert, Dahn \& Monet 1988).
The largest samples to date used to determine the white dwarf luminosity function are those of
Fleming et al. (1986; see Liebert, Bergeron \& Holberg 2004 for an updated version) on the hot
end and Oswalt et al. (1996, using white dwarfs in common proper motion binaries) on the cool end.
The most commonly used luminosity function for cool white dwarfs
(Liebert, Dahn, \& Monet 1988) was based on a sample of only 43 stars
selected on the basis of large proper motion
from the Luyten Half Second Proper Motion Survey (Luyten 1979).
Questions about completeness
and kinematic selection bias have been raised over the years, and the
need to construct a larger, deeper and more complete sample has been
obvious. Of even greater interest is the possibility of delineating a
useful sample of white dwarfs from the local halo, which may be drawn
largely from a single burst of star formation at a greater age than
the disk. Oppenheimer et al. (2001) claimed to have found a significant
population of halo white dwarfs from kinematic surveys, though
these claims were later disputed by several investigators (Reid et al. 2001; Reyle
et al. 2001; Silvestri, Oswalt, \& Hawley 2002; Bergeron 2003; Spagna et al. 2004).

The Sloan Digital Sky Survey (SDSS, York et al. 2000) offers a valuable new resource which
may be used to identify a significantly larger white dwarf sample. Imaging is performed in five
broad optical bands ($u, g, r, i,$ and $z$) down to $\sim$ 22 magnitude in $u$, $g$, and
$r$ with 95\% completeness for point sources.
Hot white dwarfs can be identified efficiently due to their blue colors
(Fan 1999; Kleinman et al. 2004), but white dwarfs
near or below the temperatures of Population II main sequence turnoff stars are
buried in the stellar locus. Methods such as the use of an intermediate-band filter
to find stars with no MgH absorption feature (Claver 1995) turned out to be
less efficient than expected (Kilic et al. 2004). 
However, the reduced proper motion technique (Luyten 1918)
offers an efficient means to identify cooler white dwarfs,
as well as halo white dwarfs, by their underluminosity in comparison to main sequence
stars with similar colors, and their high space motions.

An improved proper-motion catalog combining the USNO-B (Monet et al. 2003; 5 epochs) and SDSS catalogs
in the area of sky covered by SDSS Data Release 1 (DR1; Abazajian et al. 2003) is presented by Munn et al. (2004).
They used SDSS astrometry to recalibrate the USNO-B plate astrometry,
reducing both the statistical and systematic errors significantly.
In addition, SDSS positions were used to eliminate the large number of false high proper motion objects in the USNO-B catalog.

The combination of accurate SDSS photometry and SDSS+USNO-B astrometry enables us
to construct a reduced proper motion diagram and select cool white dwarf candidates from the disk and halo.
This paper presents results from the first year of our spectroscopic campaign.
In \S 2 we present the reduced proper motion diagram from SDSS Data Release 2 and
review our target selection criteria for follow-up spectroscopy. Our
spectroscopic observations are described in \S 3,
while an analysis of the observational
material and results from this analysis is presented in \S 4. 
Various implications of these results are then discussed in \S 5.

\section{Target Selection}

The reduced proper motion, defined as H = $m$ + 5 log$\mu$ + 5, where $m$ is the
apparent magnitude and $\mu$ is the proper motion in arcseconds per year, has long
been used as a proxy for the absolute magnitude of a star, for a sample with similar
kinematics.
Munn et al. (2004) presented a reduced proper motion diagram for a portion of SDSS DR1
in their Figure 12. 
Their SDSS+USNO-B catalog is 90\% complete to $g=19.7$, with proper motion errors
$\sim$ 3.5 mas yr$^{-1}$ in right ascension and declination. 
Four populations are delineated as roughly parallel
diagonal distributions in their diagram.  The old Population I main sequence is
seen in the top right, and the Population
II main sequence separates fairly cleanly to the left and extends down
past $H_g$ = 21 near $g-i=2$.  The white dwarf sequence
appears to separate to the left of the Population II
subdwarfs, and an unconfirmed extension of this appears as a sequence of
objects to higher $H_g$ and redder color.

Figure 1 presents the reduced proper motion diagram for all stars in SDSS Data
Release 2 (DR2; Abazajian et al. 2004) with $15 < g < 20$ and with reliably measured proper motions
greater than 20 mas yr$^{-1}$.  Individual stars are plotted only if
they are bluer than $g-i = 0$ or if $H_g > 15.136 + 2.727 (g-i)$, a cut which
should include all white dwarfs;  the remaining vast majority of stars are
represented by the contours, as there are too many stars to plot individually.
Spectroscopically confirmed white dwarfs, white dwarf plus M dwarf binaries,
subdwarfs, and QSOs are plotted as blue triangles, green triangles,
red squares, and cyan circles, respectively.  The spectroscopic
identifications are drawn from the SDSS Data Release 1 white dwarf catalog
(Kleinmann et al. 2004) and QSO catalog (Schneider et al. 2003), and the
McCook \& Sion (2003) catalog.  We also
classified all currently available SDSS spectra for stars in the diagram
with $g-i > 0$ or $H_g > 19$, and which had not previously been classified
(190 objects total).
The expected sequences of
white dwarfs (pure H atmosphere, log g = 8) for specific tangential velocities ($V_{\rm T}$) are shown as solid lines,
where colors and absolute magnitudes are predicted using model atmospheres from
P. Bergeron (private communication).
The $V_{\rm T}=$ 20--40 km s$^{-1}$ curves mark the expected location of disk white dwarfs, whereas the $V_{\rm T}=$ 150 km s$^{-1}$ curve represents the halo white dwarfs.
The white dwarf cooling curves pass through the
locus of hotter white dwarfs, then make a sharp turn due to the
onset of collision-induced absorption (CIA) due to molecular hydrogen
for the coolest stars with pure H atmospheres (Hansen 1998; Saumon \& Jacobson 1999); this opacity depresses the
$i$ band, making the colors turn bluer.

Defining a ``reliable" proper motion is important, as even a small fraction
of stars with falsely measured large proper motions will scatter stars from
the very densely populated subdwarf turnoff region of the reduced proper motion
diagram into the sparsely populated region expected to contain the cool white
dwarfs.  We
adopt a prescription for a reliably measured proper motion similar to that
delineated by Munn et al. (2004) for their catalog: (1) the SDSS detection
in both the $g$ and $i$ band must match the ``clean" criteria as described in
the DR1 documentation, (2) there must be a one-to-one match between the SDSS
and USNO-B objects, and (3) the proper motion fit must have an rms residual
of less than 525 mas in each coordinate.  We further require the star to have
been detected in all 5 epochs in USNO-B, and for the distance to the nearest
neighbor with $g < 22$ to exceed 7 arcsec.
The final two requirements are
based on inspection of all the plate images used in the USNO-B catalog for a
subsample of 562 stars in the portion of the reduced proper motion diagram
expected to be populated by halo and cool disk white dwarfs
($-1 < g < 1.5$ and $H_g > 18$), and with proper motions
greater than 100 mas yr $^{-1}$ (the reality of measured proper motions smaller
than that were too difficult to judge using the quick by-eye inspections we
employed).  Of the 201 inspected stars that were detected in all 5 USNO-B
epochs, and whose nearest neighbor (brighter than $g = 22$) is greater than
7 arcsec away, only 3 were judged to have a falsely measured large proper
motion, for a 1.5\% contamination rate.  On the other hand, of the 20
inspected stars that were detected in all 5 USNO-B epochs but had a neighbor
closer than
7 arcsec, 7 were judged to have an incorrectly measured proper motion, for
a contamination rate of 35\%.  Objects separated by less than 7 arcsec tend to
be blended on the Schmidt plates, leading to incorrectly measured proper
motions.  All of the 208 inspected stars which were not detected in one or two
of the USNO-B epochs and which have a nearest neighbor closer than 7 arcsec had
incorrectly measured proper motions.  Even if the nearest neighbor is greater
than 7 arcsec away, the contamination rate of falsely measured large proper
motions increases for stars not detected in all 5 USNO-B epochs.  For such
stars, 20 of 39 inspected stars not detected in one of the USNO-B epochs had
incorrectly measured proper motions, for a contamination rate of 51\%, while
81 of 91 not detected in two of the USNO-B epochs had incorrectly
measured proper motions, for a contamination rate of 89\%.
                                                                                                
In Figure 1, it is clear that the hot white dwarfs previously targetted
spectroscopically by SDSS are well separated from main sequence and subdwarf
stars (represented by the contours) and are located where expected from the
models.  It is also clear that the white dwarf locus extends redder and to
larger reduced proper motion, well separated from the subdwarf locus
(the diagonal cut where we start to plot individual stars), and that SDSS
has not spectroscopically observed these cooler white dwarf candidates.
This drove our target selection.
Our goal was to obtain spectroscopic identifications for a large sample of
cool white dwarf candidates so as to understand the efficiency of using
reduced proper motions to select cool white dwarfs as well as the
contamination due to subdwarfs.
Thus we selected most of our targets from the region with $g-i > 0$ and
to the left of the $V_T = 20$ km s$^{-1}$ curve, though we also selected some
to the right of the curve to better understand how cleanly the white dwarf and
subdwarf loci separate.

\section{Observations}

Follow-up spectroscopy of the cool white dwarf candidates
were obtained at the HET, the MMT, and the McDonald 2.7m Harlan Smith Telescope between
September 2003 and October 2004 (an additional observing run with the Kitt Peak 4m Telescope was completely
lost to weather).
We used the HET equipped with the Marcario Low Resolution Spectrograph (LRS) to obtain low
resolution spectroscopy of 22 cool white dwarf
candidates. Grism 2 with a 1.5\arcsec\ slit produced spectra with a resolution of 6 \AA\ over the
range 4280 -- 7340 \AA. Spectroscopy for 56 additional stars was obtained at the MMT
with the Blue Channel Spectrograph and the 500 {\it l}/mm grating, which produced spectra with a
resolution of 3.6 \AA\ over the range 3640 -- 6800 \AA.
In addition, we obtained spectroscopy for 89 stars at the McDonald 2.7m Telescope with the Large Cassegrain Spectrograph (LCS) and TI1 camera using
grating No. 43 (600 {\it l}/mm), which produced spectra with a resolution of 5.2 \AA\ over the range 3870 --
5260 \AA. In each case, a spectrophotometric standard star was observed each night for flux calibration.
Ne--Cd, He--Ar--Ne, and Ar calibration lamp exposures were taken after each observation
with the HET, the MMT, and the McDonald 2.7m, respectively.
The data were reduced using standard IRAF\footnote{IRAF is distributed by the National Optical Astronomy Observatory, which is operated by the Association of Universities for Research in Astronomy (AURA), Inc., under cooperative agreement with the National Science Foundation.} routines.

The spectra for all of the observed white dwarfs at the HET, the MMT, and the McDonald 2.7m are shown in Figures 2, 3, and 4, respectively. The spectra are ordered in decreasing $g-r$ color for Figures 2 and 3, and decreasing
$u-r$ color for Figure 4. The
majority of the objects observed at the HET and the MMT are featureless cool DC white dwarfs. The
only features seen in these spectra are due to sky subtraction problems at 5577, 5890/5896, 6300,
and the atmospheric $B$ band at 6890 \AA. Several white dwarfs show only a weak H$\alpha$
line and are DA white dwarfs with probable cool, H-rich atmospheres. Most of the brighter objects
observed at the McDonald 2.7m are DA white dwarfs. Three objects clearly show additonal features due
to late type star companions (J0929+5547, J1336+0017, J2340-1106). Two objects observed at the
2.7m are magnetic DAs (J1011+0029, J1144+6629), while two other objects observed at the 2.7m
and an additional three objects observed at the MMT are DZAs with detectable Ca II H and K lines and
Balmer lines (J0045+1420, J0748+3506, J1627+4859, J1654+3829, J2154+1300). 
We also note that four previously known white dwarfs, namely SDSS J0109-1042 (LP707-8), J0755+3621 (WD0752+365),
J1025+0043 (LHS 282), and J1145+6305 (WD1143+633) were observed at the MMT and the McDonald 2.7m, and are
included in Figures 3 and 4.

The biggest difficulty in classifying the spectra is to distinguish DZ, DZA and DAZ white dwarfs\footnote{Note that the
difference between DAZ and DZA stars is whether the dominant atmospheric constituent is hydrogen or helium,
respectively. DZA stars show steep Balmer decrements at higher $T_{\rm eff}$ than DAZ stars.}
with refractory heavy elements from nondegenerate probable-main sequence
stars (subdwarf F, G, and K types) with low, scaled-solar heavy elements.
Ca appears most frequently in DZ, DAZ and DZA stars, followed by Mg, Fe and
occasionally Na.  The white dwarfs apparently never show the CH band
(4300 \AA) which is detected for sdG and cooler stars ($g-i>0.4$), and do not show MgH and CaH
which appear in progressively-cooler sdK stars.

Figure 5 shows the spectra for several subdwarfs observed at the MMT. The spectra are plotted
in order of decreasing $T_{\rm eff}$, represented by increasing $g-i$ color. 
This figure demonstrates that subdwarfs have Ca II (hence need MMT and McDonald 2.7m blue coverage),
and usually many other metal features plus the CH molecular band. Additional MgH and CaH bands can be seen among cooler subdwarfs
(without enough blue coverage, HET was mostly used to observe cool white dwarf candidates with similar
colors to G and K type subdwarfs).
We paid close attention to the $u-g$ and $g-r$ colors of each star,
comparing colors of each candidate with white dwarf model colors
to see if they were consistent with the strengths of the H lines, and
the degree to which cool DA Balmer decrements steepen until for the
coolest DA's only H$\alpha$ is seen.

Our classifications and additional data for the 112 spectroscopically-confirmed white dwarfs and 55 subdwarfs from this study are given in Tables 1 and 2, respectively.
Positions are those from the SDSS Astrometric pipeline (Pier et al. 2003). The photometric
calibration is based on the SDSS standard star system (Smith et al. 2002) tied to the survey data
with the Photometric Telescope (Hogg et al. 2001). Interstellar absorption $A_u$ from
Schlegel, Finkbeiner \& Davis (1998), and fully dereddened magnitudes and colors are given for each object. 
Tables 1 and 2 also give the proper motions
in each component and the number of epochs in which an object is detected.
In addition, the effective temperatures, photometric distances and the tangential velocities for the newly
discovered cool white dwarfs (see \S 4.4) are also given in Table 1.

A few stars in Table 1 were previously known as likely white dwarfs. The most certain case is
SDSS J131313.12+022645.8 (LHS2696) which has a (preliminary) parallax in Dahn et al. (1989).
There are 13 additional stars discovered in the NLTT catalog,
(Luyten 1979), most with Luyten color class a, f, or g, that would have been considered to be nearly certain
white dwarfs on the basis of the NLTT data alone; the spectra in this paper now confirm these classifications.
The one very cool white dwarf in Table 1 that is also in the NLTT catalog is SDSS J075313.28+423001.6 (LP207-50).
It has color class m in the NLTT, so would not be considered as a white dwarf from the NLTT data alone.
In addition, three stars in Table 1 were identified from their proper motion by L\'epine et al. (2003):
SDSS J100225.85+610858.1 (LSR 1002+6108), SDSS J110731.38+485523.0 (LSR 1107+4855), and
SDSS J222233.90+122143.0 (LSR 2222+1221). The first two were classified as probably white dwarfs 
and the third as probably a subdwarf on the basis of their reduced proper motion and photographic colors.

\section{Results}
\subsection{Reduced Proper Motion Diagram}

Figure 6 repeats the reduced proper motion diagram of Figure 1, except that
the spectroscopic identifications presented in this paper are indicated rather
than previously known identifications.
Again, this is limited to
stars with reliable proper motions; that is, stars detected in all 5 USNO-B
epochs and with no neighbor brighter than $g = 22$ within 7 arcsec.
There is a clean separation between the white dwarfs and subdwarfs.
Of the 95 spectroscopically confirmed stars
bluer than $g-i = 1.5$ and below the $V_T = 20$ km s$^{-1}$ curve,
corresponding to the region we expect to find only white dwarfs,
91 are are certain white dwarfs, 2 are
certain subdwarfs, and two are probable white dwarfs for which a classification
of subdwarf can not be ruled out.  Visual inspection of the plate images
reveals that the two subdwarfs had falsely measured large proper motions,
consistent with the contamination rate derived earlier for our adopted
reliable proper motion criteria, while the two probable white dwarfs had
correctly measured proper motions.

Figure 7 shows the MMT spectra for these 2 probable white dwarfs. 
SDSS J2236+1419 ($H_{\rm g}=18.75$ and $g-i=0.52$) shows Mg I 3830, Ca II 3933, 3968, Ca I 4226,
Na I 5892 and probable Mg I 5175 with no detectable H$\alpha$, H$\beta$, and CH band.
The absent H and CH features suggest a very non-Population II (scaled solar) metallicity DZ white dwarf.
SDSS J1214+6216 ($H_{\rm g}=21.97$ and $g-i=0.64$) shows Mg I 3830, Ca II, no Ca I, no CH, weak H-beta,
possible Na I 5892, and rather strong H-alpha.
The steep H decrement is consistent with a cool white dwarf
with pressure-broadening, quenching the higher Balmer levels.
Again, the absent CH is probably inconsistent with an sdG.
We classify this object as a probable DZA white dwarf, pending detailed analysis.
Assuming that these two objects are white dwarfs, we find no contamination from subdwarfs with
correctly measured proper motions. On the other hand, if we assume that these objects
are in fact subdwarfs, they would represent subdwarfs that truly have scattered into the cool white dwarf region of the
reduced proper motion diagram; there is still potential for some genuine contamination from subdwarfs.

We also observed stars that did not meet our criteria for a reliable
proper motion, having either not been detected in 1 or 2 USNO-B epochs or
having a neighbor within 7 arcsec. Of these, 16 were white dwarfs
(plotted as blue asterisks in Figure 6), all located within the white dwarf
region of the reduced proper motion diagram; upon visual inspection of the
plates used in USNO-B, all of the measured proper motions were correct.
Another 37 were subdwarfs (see Table 2), located both in the white dwarf and subdwarf regions
of the reduced proper motion diagram; visual inspection shows all but three of
these to have incorrectly measured proper motions, and the proper motions of
the remaining three were too small to determine their validity by eye.
                                                                                                
The reduced proper motion diagram, using our conservative criteria for a
reliable proper motion, can thus be used to define a statistically complete
sample of white dwarfs, including the coolest white dwarfs which are difficult
to efficiently select using other techniques.  There is a
roughly 1.5\% contamination rate of incorrectly measured proper motions.
There is likely no contamination due to subdwarfs with
correctly measured proper motions, though contamination of a few percent
is still possible. True white dwarfs
that fail to meet the proper motion criteria must be accounted for, either
statistically or by using visual inspection to verify proper motions for all
candidates.  This result is used in Harris et al. (2005) to construct the
white dwarf luminosity function using the SDSS Data Release 3 imaging and
USNO-B astrometry.

\subsection{Color-Color Diagrams}

SDSS color-color diagrams for spectroscopically identified proper motion objects are shown
in Figure 8. The spectral classifications are indicated in the diagrams, and contours that
show the colors of nondegenerate stars in the SDSS are included for comparison.
The curves show the colors of white dwarf model atmospheres
of pure H (solid curves) and pure He (dashed curves)
composition with log $g=$ 7, 8, and 9, kindly made available to us by P. Bergeron.

Harris et al. (2003) showed that the sequence of white dwarfs hotter than 12,000 K is
contaminated only by hot subdwarfs.
Our $u-g$ vs. $g-r$ color-color diagram demonstrates that white dwarfs and
subdwarfs selected from the reduced proper motion diagram separate from each other
for $u-g\leq$ 1.0, though this result depends on a handful of white dwarfs observed between
the range 0.6 $\leq u-g \leq$ 1.0 and further observations are needed to confirm this result.

Figure 8a also shows that the cool white dwarf model atmospheres cannot predict the $u-g$ colors
below 6,000 K accurately, though the $g-r$, $r-i$, and $i-z$ colors agree reasonably well with the
observed sequence of cool white dwarfs (see Figures 8b and 8c). Bergeron, Leggett, \& Ruiz (1997) were the first to
introduce a
UV opacity source in the coolest hydrogen-rich white dwarf models in terms of a pseudo continuum
opacity to fit the observed hydrogen-rich cool white dwarf sequence. They found that this
opacity source is
needed to explain the observed UV colors of cool hydrogen-rich white dwarfs below 5,300 K.
This is also seen in the $B-V$ vs. $V-K$ color-color diagram of Bergeron, Ruiz, \& Leggett
(2001), and $r-DDO51$ vs. $r-z$ color-color diagram of Kilic et al. (2004).
Wolff, Koester, \& Liebert (2002) showed that this UV flux deficiency (relative to model
atmosphere predictions) extends even more strongly
into the space ultraviolet region (2000 -- 3200 \AA).
Here, we find that the
$u-g$ colors for neither cool DA nor cool DC white dwarfs can be explained
with the current model atmospheres. Significant improvements are needed in the
cool white dwarf model atmospheres to understand the unexplained UV opacity
which is crucial for the coolest
white dwarfs. In fact, this may be the reason why efforts to fit the
spectral energy distributions of ultracool white dwarfs fail. 

The majority of the cool white dwarfs with $r-i>0.25$ ($T_{\rm eff}<5000 K$) tend to have bluer
$i-z$ colors compared to the model predictions (see Figure 8c); 
the observed blue turn-off of cool white dwarfs is at a
bluer color than expected from the models. All but four of the cool white dwarfs occupy a
region to the left of the log $g=9$ (left-most line) white dwarf models. This implies that either
all of the very cool white dwarfs are massive, they have mixed H/He atmospheres, or our understanding
of the CIA opacities are incomplete.
All being massive is statistically unlikely as the average mass for the cool white dwarfs ($T_{\rm eff}\leq5000 K$)
with trigonometric parallax measurements is 0.61 $\pm$ 0.2 M$_{\odot}$ (Bergeron, Leggett, \& Ruiz 2001).
Mixed atmosphere white dwarfs are expected to show stronger flux deficits in the infrared than pure H
white dwarfs (Bergeron, Saumon, \& Wesemael 1995); the effects of CIA becomes
significant at warmer temperatures, which could explain the observed blue turn-off of cool white
dwarfs in Figure 8c. Bergeron \& Leggett (2002) argued that all white dwarfs
cooler than
4000 K have mixed H/He atmospheres. In addition, Kilic et al. (2004) suggested that all white
dwarfs cooler than 5000 K may have mixed atmospheres. Therefore, this figure presents further evidence
that the coolest white dwarfs indeed have helium rich atmospheres. 

\subsection{Non-DA Gap}

The H$\alpha$ and H$\beta$ equivalent widths of all DA and DC white dwarfs in our sample
are shown against $g-r$ in Figure 9. White dwarf models predict H$\alpha$ to disappear
around $V-I \sim$ 1.1 (see Figure 7 of Bergeron, Ruiz, \& Leggett 1997).
Although there is a large scatter in our equivalent width measurements in Figure 9,
it is apparent that the H$\alpha$ and H$\beta$ equivalent widths decrease with increasing $g-r$ color and
they vanish around $g-r \sim$ 0.7 (or $V-I\sim$ 1.1), in good agreement with the model predictions.
The scatter in the equivalent width measurements may partly be due to variations in gravities and the use of
spectra with different signal-to-noise ratios from four different telescope + instrument combinations.

Bergeron, Ruiz \& Leggett (1997) and Bergeron, Leggett \& Ruiz (2001) argued that
most of the white dwarfs with $T_{\rm eff}$ in the range 6000 K -- 5000 K are of DA type, and they
found evidence for a so-called non-DA gap. Even though non-DA stars are seen above and below this temperature
range, they found only two peculiar non-DA stars inside the gap. Bergeron, Ruiz \& Leggett (1997),
Bergeron, Leggett \& Ruiz (2001), and Hansen (1999) tried to explain the existence of this gap in terms of
different physical mechanisms, i.e. convective mixing and different evolutionary timescales of hydrogen-rich
and helium-rich white dwarfs. The existence of the non-DA gap is apparent in Bergeron, Leggett \& Ruiz's (2001)
$V-I$ vs. $V-K$ color-color diagram.
Figure 10 shows $u-g$ vs. $g-r$ (left panel), $g-r$ vs. $r-i$ (middle panel), and
$r-i$ vs. $i-z$ (right panel) color-color diagrams for white dwarfs in our sample. The data set is divided into
DA (upper panels) and non-DA (lower panels, DC and DQ) stars. The pure H (DA panels) and pure He (non-DA panels)
model sequences with $6000K\ga T_{\rm eff}\ga5000K$ are also shown for log $g=7$ (solid line), log $g=8$ (dashed line), and log $g=9$ (dotted line).
The two differences between these color-color diagrams and Figure 8 are that (1) DC white dwarfs discovered at the McDonald
2.7m are not included in Figure 10 since the 2.7m spectra do not cover H$\alpha$, and so some of the white
dwarfs observed with the 2.7m and classified as DCs may turn out to be DAs showing H$\alpha$ only, and
(2) ultra-cool white dwarfs discovered in the SDSS are not included. SDSS J1001+3903, an ultra-cool
white dwarf discovered by Gates et al. (2004), falls in the
non-DA gap in the $u-g$ vs. $g-r$ diagram, but not in the other two diagrams.
Even though ultra-cool white 
dwarfs are predicted to have $T_{\rm eff}\leq4000 K$, they mimic bluer/warmer objects in the
various color-color diagrams and therefore are not included in this discussion.

DA white dwarfs show H$\alpha$ absorption for $g-r\leq0.7$ ($\geq 5000K$) and we should have detected H$\alpha$
for DCs with $6000K\ga T_{\rm eff}\ga5000K$ if they had pure H atmospheres.
The H$\alpha$ panel in Figure 9 shows exactly 3 DC stars in the gap (The gap, marked by the dashed lines, is $g-r$ = 0.39 to 0.64 for hydrogen
atmospheres). 
Based on $ugriz$ photometry (Figure 10), we see DC stars above and below the non-DA gap, but we find only 3 to 5 DC stars
(shown as filled red circles in lower panels) in the gap.
All of these white dwarfs (SDSS J0157+1335, J1203+0426, J1205+0449, J1648+3939, J1722+5752) are observed at the MMT and HET,
and Balmer lines cannot be seen in their spectra. Assuming pure He composition, we estimate their temperatures to be
$5254K, 5144K, 5635K, 5536K$, and $5555K$ (see the next section for temperature estimates).
These stars appear to be non-DA stars in the 5000-6000K
temperature range, and fill in the Bergeron et al non-DA gap
to some extent. The fraction of DA/non-DA stars in this
temperature range is still seen to be large in our data,
and further data are needed to quantify the ratio.
Moreover, Bergeron, Ruiz \& Leggett (1997) have identified a group of DC white dwarfs which lie close
to $T_{\rm eff}=6000K$ whose energy distributions are better reproduced with pure hydrogen models.
Therefore, some or all of these stars may be explained with H-rich compositions.
Bergeron, Leggett, \& Ruiz (2001) and Ruiz \& Bergeron (2001) showed that
infrared photometry is needed to discriminate between between hydrogen-rich
and helium-rich atmospheres for cool white dwarfs; infrared photometry is needed
to reliably determine the spectral types and temperatures for these stars.
In any case, our data supports at least a deficit in the number of non-DA white dwarfs in the predicted non-DA gap.
 
\subsection{Model Atmosphere Analysis}
The $u,g,r,i$, and $z$ photometry for each
of the new spectroscopically confirmed white dwarfs has been fitted with synthetic photometry
predicted from the model atmospheres (P. Bergeron, private communication),
using a minimum chi square fit. Since trigonometric parallax measurements are not
available for these white dwarfs, a value of log $g=$ 8.0 has been assumed for all
objects. Due to larger uncertainties in $u$ and $z$, and the fact that the $u$
magnitudes of cool white dwarfs are affected by an unexplained UV opacity source, $u$
and $z$ magnitudes are given lower weight in our fits.
The fits have been done using only $u,g$, and $r$ magnitudes for white dwarf +
late type star binaries, though some of these binaries may have contaminated
$u,g$, and $r$ magnitudes.

While spectra of bluer
white dwarfs allow us to determine whether the atmosphere is hydrogen-rich or
helium-rich, H$\alpha$ and H$\beta$ disappear around $g-r \sim 0.7$ and so IR
photometry is need to determine the atmospheric composition for DCs.
We assume hydrogen-rich composition for the analysis of all of the DA and DC white dwarfs
in our sample. Figure 11 shows the effects of using hydrogen-rich versus helium-rich composition models
on our temperature estimates for the DC stars in our sample.
Both hydrogen-rich and helium-rich models give similar answers for stars with
$T_{\rm eff}$ in the range 10000 -- 5500 K ($g-i<0.7$), though the effect drastically increases below 5500K due
to the onset of collision induced absorption. Pure He models predict warmer temperatures than the pure H models;
the optical colors of a 4000 K pure H atmosphere white dwarf can be fit with a 4700 K pure He white dwarf model.

We assumed zero reddening for white dwarfs with estimated distances $\leq$ 100 pc,
and used the full reddening value from Schlegel, Finkbeiner, \& Davis (1998) if
the estimated distance from the Galactic plane is larger than 250 pc. For white
dwarfs with
estimated distances between 100 and 250 pc, we used a linear interpolation between
zero and the full reddening coefficient.

Results of the model atmosphere fits are summarized in columns 12--16 of Table 1.
For each white dwarf, we give (assuming pure H composition with log $g=$ 8.0) the
effective temperature, the predicted bolometric magnitude, the absolute magnitude in
$g$, the distance, and the estimated tangential velocity.
According to our fits, there are seven
white dwarfs with $T_{\rm eff}\leq$ 4,000 K in our sample, but
these temperature estimates are questionable, and IR photometry is needed to obtain reliable
results for these objects. 
Although most of the newly found white dwarfs show disk
kinematics, there are 16 objects with $V_{\rm T}\geq$ 150 km s$^{-1}$
that may be halo white dwarfs. Halo membership of these objects and
several others from the SDSS Data Release 3 are discussed in Harris et al. (2005).

\section{Conclusions and Future Work}

SDSS photometry and SDSS+USNO-B astrometry (Munn et al. 2004) has been
used to isolate cool white dwarf candidates in a reduced proper motion diagram
in the SDSS Data Release 2.  Using SDSS spectra and follow-up spectroscopy
on the MMT, HET, and the McDonald 2.7m, we showed that the white dwarf locus
in the reduced proper motion diagram is cleanly separated from the far more
numerous subdwarfs, with a contamination rate due to falsely measured large
proper motions of only 1---2\%.  In a companion paper, Harris et al. (2005)
use this clean separation of white dwarfs in the reduced proper motion diagram to
assemble a statistically complete sample of white dwarfs in the SDSS Data Release 3
and determine the white dwarf luminosity function. The detailed shape
of this new luminosity function will eventually make
it possible to study the cooling physics of white dwarfs in detail, e.g. neutrino cooling,
crystallization, and phase separation. These effects produce overlapping signatures
(bumps) in the white dwarf luminosity function which can be used to calibrate their
significance. The details of these constituent input physics can affect the implied ages of
cool white dwarfs below log (L/L$_{\odot}$)$\sim -$4.2 by as much as 2-3 Gyr
(Hawkins \& Hambly 1999; Montgomery et al. 1999; Salaris et al. 2000).

All of the ultracool white dwarfs discovered in SDSS (Harris et al. 2001;
Gates et al. 2004) show significant proper motion, hence they would be discovered in our proper
motion survey if not targeted by SDSS fiber spectroscopy.
We have discovered seven new cool white dwarfs with estimated temperatures below 4000 K.
Nevertheless, these objects may have temperatures above 4000 K if they have pure He atmospheres (see Figure 11).
Our spectroscopy at the HET and the MMT do not go red enough
to detect the CIA in the near IR, though
all of these objects and several others should show infrared flux deficiency
due to CIA if they have pure H or mixed H/He atmospheres.
Our SDSS DR2 proper motion catalog does not reveal any other ultracool white dwarf
candidate exhibiting strong CIA absorption. Therefore, SDSS1337+00, LHS3250 and SDSSJ0947+44 are
the only ultracool white dwarfs (within our magnitude limit) showing strong CIA absorption
in the SDSS DR2 imaging area
(3324 square degrees). Further progress in understanding the ultracool white dwarfs and estimating
reliable temperatures for our cool white dwarf sample can be achieved with the help of $JHK$
infrared photometric observations. We have begun such a program at the NASA Infrared
Telescope Facility. Trigonometric parallax measurements for these white dwarfs will also be
necessary to measure their masses.

Even though our survey has a fainter magnitude limit and a lower proper motion cut off than the
LHS survey, the faint end of the white dwarf luminosity function where most of the age sensitivity 
resides is still poorly populated. The magnitude limit of the
SDSS+USNO-B proper motion catalog is set by the photographic POSS I and POSS II plates.
Although SDSS imaging is
95\% complete down to $g=22$, the SDSS+USNO-B proper motion catalog is only 90\% complete down
to $g=$19.7. A second epoch CCD imaging survey in $r$ or $i$ to 22 mag may be necessary to find the coolest
white dwarfs in the disk and possible halo white dwarfs (von Hippel et al. 2004). Also, we may be
missing slowly moving
cool white dwarfs; a kinematically unbiased control sample, e.g. DDO51 photometry + multi object
spectroscopy (Kilic et al. 2004), would be useful to check the completeness of our survey.

\acknowledgements
This material is based upon work supported by the National Science Foundation under Grants AST--0307315
(to TvH, DEW, and MK) and AST--0307321 (to JL and KAW).
We thank Pierre Bergeron and Didier Saumon for making their model atmospheres available to us.
We also thank Harry Shipman for careful reading of this manuscript.
The Hobby-Eberly Telescope (HET) is a joint project of the University of Texas at Austin, the Pennsylvania State University, Stanford University, Ludwig-Maximilians-Universit\"at M\"unchen, and Georg-August-Universit\"at G\"ottingen. The HET is named in honor of its principal benefactors, William P. Hobby and Robert E. Eberly. The Marcario Low Resolution Spectrograph is named for Mike Marcario of High Lonesome Optics who fabricated several optics for the instrument but died before its completion. The LRS is a joint project of the Hobby-Eberly Telescope partnership and the Instituto de Astronomía de la Universidad Nacional Autonoma de M\'{e}xico.

Funding for the Sloan Digital Sky Survey (SDSS) has been provided by the Alfred P. Sloan Foundation, the Participating Institutions, the National Aeronautics and Space Administration, the National Science Foundation, the U.S. Department of Energy, the Japanese Monbukagakusho, and the Max Planck Society. The SDSS Web site is http://www.sdss.org/.

The SDSS is managed by the Astrophysical Research Consortium (ARC) for the Participating Institutions. The Participating Institutions are The University of Chicago, Fermilab, the Institute for Advanced Study, the Japan Participation Group, The Johns Hopkins University, the Korean Scientist Group, Los Alamos National Laboratory, the Max-Planck-Institute for Astronomy (MPIA), the Max-Planck-Institute for Astrophysics (MPA), New Mexico State University, University of Pittsburgh, Princeton University, the United States Naval Observatory, and the University of Washington.

\clearpage
\begin{deluxetable}{lcrrrrrrrrlrrrrrr}
\tabletypesize{\tiny}
\rotate
\tablecolumns{17}
\tablewidth{0pt}
\tablecaption{Spectroscopically Identified White Dwarfs}
\tablehead{
\colhead{Name (SDSS J)}&
\colhead{$g$}&
\colhead{$u-g$}&
\colhead{$g-r$}&
\colhead{$r-i$}&
\colhead{$i-z$}&
\colhead{$A_{\rm u}$}&
\colhead{$\mu_{\rm ra}$}&
\colhead{$\mu_{\rm dec}$}&
\colhead{Ep}&
\colhead{Type}&
\colhead{$T_{\rm eff}$}&
\colhead{$M_{\rm bol}$}&
\colhead{$M_{\rm g}$}&
\colhead{D}&
\colhead{$V_{\rm T}$}&
\colhead{Source}
}
\startdata
00 03 16.69$-$01 11 17.9& 19.21& 1.22& 0.49& 0.22& 0.02& 0.16& 98& $-$16& 6& DA& 5351& 14.57& 15.14& 69.19& 32.57& MMT\\
00 11 42.67$-$09 03 24.3& 17.73& 0.64& 0.31& 0.12& 0.01& 0.21& 4& $-$134& 6& DA& 6125& 13.97& 14.37& 50.31& 31.97& 2.7m\\
00 28 37.06$-$00 29 28.9& 19.68& 0.54& 0.36& 0.04& $-$0.11& 0.10& 103& 67& 6& DC& 6381& 13.80& 14.16& 128.79& 75.01& HET\\
00 43 16.02$+$15 40 59.5& 18.11& 0.53& 0.18& 0.01& $-$0.05& 0.28& $-$37& 25& 6& DA& 6844& 13.49& 13.82& 79.16& 16.76& 2.7m\\
00 45 21.88$+$14 20 45.3& 18.81& 1.35& 0.63& 0.21& 0.03& 0.55& 260& $-$53& 6& DZA& 4732& 15.11& 15.90& 45.73& 57.51& MMT\\
01 02 59.98$+$14 01 08.1& 19.29& 1.71& 0.73& 0.27& 0.11& 0.22& 12& 106& 6& DC& 4582& 15.25& 16.10& 46.84& 23.68& MMT\\
01 15 14.73$+$14 35 57.5& 18.54& 0.52& 0.25& 0.09& $-$0.02& 0.31& $-$45& $-$55& 4& DA& 6320& 13.84& 14.21& 81.60& 27.49& MMT\\
01 28 27.47$-$00 45 12.6& 17.74& 1.02& 0.37& 0.15& 0.00& 0.16& 147& $-$43& 6& DA& 5854& 14.18& 14.62& 44.54& 32.33& 2.7m\\
01 57 43.25$+$13 35 58.2& 19.15& 1.19& 0.64& 0.16& 0.05& 0.25& 87& $-$62& 6& DC& 5040& 14.84& 15.51& 57.63& 29.18& MMT\\
01 59 38.43$-$08 12 42.4& 19.80& 0.42& 0.09& $-$0.10& 0.01& 0.13& 322& $-$119& 6& DA& 8214& 12.69& 13.01& 230.40& 374.90& MMT\\
02 50 05.81$-$09 10 02.8& 18.87& 1.06& 0.48& 0.18& 0.05& 0.15& 106& DA& 6& DA& 5474& 14.47& 15.00& 62.59& 31.45& MMT\\
02 56 41.62$-$07 00 33.8& 18.81& 1.73& 0.80& 0.34& 0.08& 0.26& 373& $-$202& 6& DC& 4211& 15.62& 16.56& 30.94& 62.22& MMT\\
02 58 54.42$+$00 30 40.4& 18.84& 0.73& 0.27& 0.06& $-$0.07& 0.42& $-$80& 16& 6& DA& 6189& 13.93& 14.32& 92.32& 35.70& MMT\\
03 09 24.87$+$00 25 25.3& 17.75& 0.84& 0.34& 0.13& $-$0.02& 0.57& $-$6& $-$106& 6& DC& 5637& 14.34& 14.83& 46.60& 23.45& 2.7m\\
03 14 49.81$-$01 05 19.3& 18.31& 0.88& 0.38& 0.12& $-$0.04& 0.39& $-$77& $-$71& 6& DA& 5709& 14.29& 14.76& 58.48& 29.03& 2.7m\\
03 16 13.90$-$08 16 37.6& 16.63& 0.56& 0.17& 0.02& $-$0.04& 0.48& 90& $-$103& 6& DA& 6610& 13.64& 13.99& 39.65& 25.71& 2.7m\\
03 30 54.88$+$00 37 16.5& 19.33& 0.81& 0.34& 0.10& 0.01& 0.57& 77& 34& 6& DA& 5690& 14.30& 14.78& 98.78& 39.41& MMT\\
04 06 32.39$-$04 32 50.4& 17.02& 0.53& 0.17& 0.00& $-$0.10& 0.56& 171& 80& 6& DA& 6624& 13.63& 13.98& 48.89& 43.75& 2.7m\\
04 06 47.32$-$06 44 36.9& 17.70& 0.74& 0.34& 0.07& 0.03& 0.45& 67& 27& 6& DA& 5884& 14.15& 14.59& 48.70& 16.67& 2.7m\\
07 48 11.90$+$35 06 32.4& 18.08& 1.13& 0.36& 0.09& $-$0.07& 0.30& $-$44& $-$141& 6& DZA& 5925& 14.12& 14.55& 56.16& 39.32& 2.7m\\
07 53 13.28$+$42 30 01.6& 17.91& 1.84& 0.84& 0.30& 0.10& 0.23& 113& $-$403& 6& DC& 4226& 15.61& 16.54& 20.36& 40.39& 2.7m\\
07 56 31.11$+$41 39 50.9& 16.76& 0.52& 0.18& 0.03& $-$0.05& 0.20& $-$9& $-$349& 6& DA& 6951& 13.42& 13.75& 42.59& 70.48& 2.7m\\
08 20 36.99$+$43 10 05.3& 17.40& 0.49& 0.08& 0.03& $-$0.09& 0.33& $-$63& $-$103& 6& DA& 7192& 13.27& 13.59& 64.72& 37.04& 2.7m\\
08 20 56.07$+$48 03 52.9& 17.16& 0.62& 0.24& 0.12& $-$0.06& 0.22& 224& $-$80& 6& DA& 6388& 13.79& 14.16& 42.91& 48.37& 2.7m\\
08 23 07.81$+$48 33 16.6& 17.79& 0.74& 0.25& 0.10& $-$0.02& 0.22& $-$217& $-$72& 6& DA& 6378& 13.80& 14.16& 57.26& 62.06& 2.7m\\
08 25 19.70$+$50 49 20.1& 19.17& 1.72& 0.86& 0.33& 0.05& 0.24& $-$331& $-$330& 6& DC& 4048& 15.79& 16.74& 33.42& 74.04& MMT\\
08 36 41.56$+$45 56 58.7& 19.89& 1.63& 0.84& 0.27& 0.15& 0.15& $-$64& $-$169& 6& DC& 4373& 15.46& 16.36& 53.19& 45.56& MMT\\
08 37 12.30$+$46 13 25.1& 18.39& 0.69& 0.29& 0.09& 0.00& 0.14& $-$80& $-$40& 6& DA& 6363& 13.81& 14.18& 72.90& 30.90& MMT\\
09 19 48.92$+$01 13 53.0& 18.21& 0.73& 0.30& 0.11& 0.00& 0.13& 137& $-$193& 6& DA& 6227& 13.90& 14.29& 63.81& 71.58& 2.7m\\
09 29 03.12$+$55 47 58.5& 17.85& 0.43& 0.23& 1.05& 0.87& 0.15& $-$350& $-$18& 6& DA+M& 6719& 13.57& 13.91& 64.55& 107.23& 2.7m\\
09 34 38.94$+$53 29 37.4& 17.47& 0.51& 0.20& 0.04& $-$0.01& 0.06& $-$151& $-$145& 6& DA& 6976& 13.40& 13.73& 57.12& 56.68& 2.7m\\
09 42 44.96$+$44 37 43.1& 19.44& 1.92& 0.88& 0.37& 0.19& 0.06& $-$135& $-$189& 6& DC& 4052& 15.79& 16.73& 35.66& 39.25& HET\\
10 01 19.48$+$46 56 50.6& 19.24& 2.06& 1.06& 0.33& 0.09& 0.07& $-$17& $-$339& 6& DC& 3284& 16.71& 17.48& 23.08& 37.13& HET\\
10 02 25.85$+$61 08 58.1& 19.34& 2.32& 0.96& 0.38& 0.18& 0.08& $-$448& $-$328& 6& DC& 3581& 16.33& 17.20& 27.79& 73.15& MMT\\
10 05 21.05$+$53 54 08.4& 18.02& 0.50& 0.24& 0.05& $-$0.02& 0.04& $-$145& $-$219& 6& DA& 6800& 13.52& 13.85& 68.91& 85.79& 2.7m\\
10 11 05.63$+$00 29 44.4& 17.23& 0.69& 0.31& 0.11& $-$0.02& 0.18& $-$219& 55& 6& DAH& 6184& 13.93& 14.32& 40.48& 43.33& 2.7m\\
10 13 59.85$+$03 05 53.8& 18.60& 1.37& 0.63& 0.23& 0.10& 0.16& 107& $-$101& 6& DA& 4964& 14.90& 15.60& 42.00& 29.29& MMT\\
10 14 14.45$+$04 01 37.4& 16.76& 0.46& 0.11& 0.01& $-$0.06& 0.11& $-$199& 26& 6& DA& 7506& 13.08& 13.40& 48.78& 46.41& 2.7m\\
10 22 10.36$+$46 12 49.2& 16.42& 0.42& 0.20& 0.05& $-$0.06& 0.07& 9& $-$121& 6& DA& 6993& 13.39& 13.72& 35.38& 20.35& 2.7m\\
10 23 56.10$+$63 48 33.8& 18.08& 0.86& 0.34& 0.09& 0.00& 0.05& $-$344& $-$216& 6& DA& 6243& 13.89& 14.28& 58.47& 112.58& 2.7m\\
10 48 01.84$+$63 34 48.9& 17.90& 1.43& 0.61& 0.29& 0.02& 0.04& $-$258& $-$142& 6& DA& 5004& 14.87& 15.55& 29.94& 41.79& MMT\\
11 02 13.70$+$67 07 52.6& 19.55& 1.74& 0.65& 0.29& 0.03& 0.09& $-$380& $-$185& 6& DC& 4840& 15.01& 15.76& 59.24& 118.67& HET\\
11 07 31.38$+$48 55 23.0& 19.39& 2.02& 0.92& 0.30& 0.12& 0.11& $-$726& $-$79& 6& DC& 4020& 15.82& 16.77& 34.85& 120.64& HET\\
11 11 54.54$+$03 37 26.2& 18.22& 0.91& 0.37& 0.11& 0.06& 0.21& $-$371& $-$127& 6& DA& 5899& 14.14& 14.57& 57.72& 107.28& 2.7m\\
11 13 06.26$+$00 32 43.7& 17.60& 0.53& 0.21& $-$0.07& $-$0.04& 0.34& $-$363& $-$89& 6& DC& 6953& 13.42& 13.75& 65.60& 116.21& 2.7m\\
11 15 36.96$+$00 33 17.3& 17.75& 1.52& 0.66& 0.24& 0.05& 0.23& 37& $-$250& 6& DA& 4816& 15.04& 15.79& 26.57& 31.83& MMT\\
11 19 40.62$-$01 07 55.1& 19.79& 2.01& 0.85& 0.24& 0.15& 0.23& $-$291& $-$28& 6& DC& 4283& 15.55& 16.47& 49.77& 68.97& HET\\
11 36 55.18$+$04 09 52.6& 16.98& 0.45& $-$0.10& 0.32& 0.51& 0.12& $-$96& $-$53& 4& DA& 10077& 11.79& 12.18& 95.28& 49.52& 2.7m\\
11 43 52.16$-$01 31 49.4& 17.36& 0.72& 0.24& 0.09& $-$0.02& 0.10& $-$277& $-$2& 6& DA& 6594& 13.65& 14.00& 48.57& 63.78& 2.7m\\
11 44 39.54$+$66 29 28.5& 17.47& 0.62& 0.22& 0.04& $-$0.06& 0.05& $-$145& $-$20& 6& DAH& 6919& 13.44& 13.77& 55.76& 38.68& 2.7m\\
11 46 25.77$-$01 36 36.9& 16.48& 0.59& 0.25& 0.12& $-$0.06& 0.07& 358& $-$434& 6& DA& 6516& 13.70& 14.05& 31.35& 83.61& 2.7m\\
12 02 00.48$-$03 13 47.4& 19.86& 2.37& 0.85& 0.32& 0.06& 0.15& $-$73& 134& 6& DC& 4151& 15.69& 16.62& 47.03& 34.01& MMT\\
12 03 28.65$+$04 26 53.4& 18.11& 1.36& 0.66& 0.28& 0.08& 0.10& $-$252& 156& 6& DC& 4852& 15.00& 15.75& 30.66& 43.07& MMT\\
12 04 39.54$+$62 22 16.4& 19.16& 1.69& 0.78& 0.29& 0.10& 0.10& $-$21& $-$159& 6& DC& 4528& 15.31& 16.17& 40.92& 31.11& MMT\\
12 05 29.15$+$04 49 35.6& 18.45& 0.89& 0.48& 0.19& 0.07& 0.09& $-$138& $-$53& 6& DC& 5524& 14.43& 14.94& 51.65& 36.19& MMT\\
12 33 22.45$+$06 07 10.7& 18.21& 1.32& 0.55& 0.19& 0.03& 0.09& $-$79& $-$352& 6& DA& 5302& 14.61& 15.19& 41.41& 70.80& 2.7m\\
12 34 08.12$+$01 09 47.4& 19.73& 1.40& 0.54& 0.25& 0.03& 0.13& $-$284& $-$55& 6& DA& 5177& 14.72& 15.34& 79.01& 108.34& HET\\
12 38 47.85$+$51 22 07.4& 17.32& 0.46& 0.12& $-$0.02& $-$0.13& 0.07& $-$319& $-$21& 6& DA& 7710& 12.96& 13.29& 65.38& 99.06& 2.7m\\
13 00 21.25$+$01 30 45.5& 17.74& 1.23& 0.54& 0.18& 0.12& 0.10& $-$374& 145& 6& DA& 5297& 14.61& 15.20& 33.48& 63.66& 2.7m\\
13 01 21.14$+$67 13 07.4& 16.69& 0.56& 0.25& 0.09& $-$0.04& 0.06& 151& 52& 4& DA& 6629& 13.63& 13.97& 35.51& 26.88& 2.7m\\
13 03 13.03$-$03 23 23.9& 16.81& 0.45& 0.16& 0.02& $-$0.09& 0.13& 32& $-$137& 6& DA& 7160& 13.29& 13.62& 45.47& 30.32& 2.7m\\
13 13 13.12$+$02 26 45.8& 18.84& 2.04& 1.07& 0.37& 0.18& 0.14& $-$744& $-$116& 6& DC& 3394& 16.56& 17.38& 20.23& 72.21& MMT\\
13 22 54.60$-$00 50 42.8& 18.82& 1.75& 0.76& 0.31& 0.09& 0.14& $-$156& 118& 6& DC& 4505& 15.33& 16.20& 35.08& 32.52& MMT\\
13 36 16.05$+$00 17 32.7& 17.34& 0.51& 0.36& 0.76& 0.58& 0.12& $-$278& $-$141& 6& DA+M& 6116& 13.98& 14.38& 40.76& 60.23& 2.7m\\
13 39 39.55$+$67 04 49.8& 19.79& 0.75& 0.28& 0.12& $-$0.02& 0.07& $-$194& 235& 5& DA& 6409& 13.78& 14.14& 137.61& 198.77& HET\\
13 40 43.35$+$02 03 48.3& 18.01& 1.11& 0.43& 0.20& 0.04& 0.13& $-$534& 28& 6& DC& 5600& 14.37& 14.87& 44.45& 112.66& 2.7m\\
13 57 58.43$+$60 28 55.3& 18.04& 0.72& 0.35& 0.11& $-$0.05& 0.07& $-$304& 55& 6& DC& 6186& 13.93& 14.32& 56.53& 82.78& 2.7m\\
14 22 25.73$+$04 59 39.7& 19.34& 1.54& 0.83& 0.30& 0.08& 0.15& $-$277& $-$62& 6& DC& 4365& 15.47& 16.37& 41.08& 55.27& HET\\
14 26 59.40$+$49 21 00.6& 16.93& 0.55& 0.18& 0.07& $-$0.09& 0.11& $-$96& 44& 6& DC& 6927& 13.43& 13.77& 44.52& 22.28& 2.7m\\
14 52 24.95$-$00 11 34.7& 18.25& 1.36& 0.58& 0.20& 0.07& 0.27& 155& 129& 6& DC& 5052& 14.83& 15.49& 38.98& 37.26& 2.7m\\
15 48 35.89$+$57 08 26.4& 17.70& 0.80& 0.39& 0.13& 0.05& 0.06& $-$220& $-$138& 6& DC& 5975& 14.08& 14.50& 44.38& 54.64& 2.7m\\
15 55 34.18$+$50 25 47.8& 16.70& 0.75& 0.31& 0.13& 0.00& 0.10& $-$234& $-$5& 6& DA& 6204& 13.92& 14.31& 31.13& 34.54& 2.7m\\
16 09 20.13$+$52 22 39.6& 18.21& 0.65& 0.25& 0.12& 0.01& 0.10& 155& 281& 6& DA& 6467& 13.73& 14.09& 68.91& 104.82& 2.7m\\
16 15 44.67$+$44 49 42.5& 19.56& 1.63& 0.75& 0.27& 0.08& 0.05& 44& $-$237& 6& DC& 4698& 15.14& 15.95& 53.41& 61.02& HET\\
16 23 24.05$+$34 36 47.7& 17.18& 0.39& 0.11& 0.08& 0.45& 0.10& $-$58& 105& 6& DA& 7650& 13.00& 13.32& 61.11& 34.74& 2.7m\\
16 27 12.99$+$00 28 18.6& 17.38& 0.78& 0.31& 0.06& $-$0.01& 0.47& $-$194& $-$73& 6& DA& 5983& 14.08& 14.49& 44.31& 43.54& 2.7m\\
16 27 31.09$+$48 59 19.0& 19.19& 1.51& 0.59& 0.24& 0.05& 0.07& $-$91& 77& 6& DZA& 5105& 14.78& 15.43& 58.01& 32.78& MMT\\
16 48 47.07$+$39 39 17.0& 18.81& 1.29& 0.54& 0.16& 0.06& 0.07& $-$126& 0& 6& DC& 5401& 14.53& 15.08& 56.98& 34.03& MMT\\
16 54 45.70$+$38 29 36.6& 16.93& 0.96& 0.40& 0.15& 0.02& 0.08& 18& $-$325& 5& DZA& 5847& 14.18& 14.62& 29.72& 45.85& 2.7m\\
16 59 40.00$+$32 03 20.1& 17.56& 0.63& 0.28& 0.07& 0.01& 0.16& $-$238& $-$244& 6& DA& 6428& 13.76& 14.12& 51.35& 82.96& 2.7m\\
17 04 47.70$+$36 08 47.4& 18.63& 1.79& 0.75& 0.28& 0.12& 0.13& 186& $-$175& 6& DC& 4560& 15.28& 16.13& 33.21& 40.20& HET\\
17 14 33.26$+$27 38 36.1& 18.16& 0.40& 0.12& 0.01& $-$0.11& 0.24& 49& $-$20& 6& DC& 7235& 13.24& 13.57& 89.96& 22.57& 2.7m\\
17 22 57.78$+$57 52 50.7& 19.17& 1.14& 0.46& 0.23& 0.06& 0.15& $-$37& 390& 6& DC& 5403& 14.53& 15.08& 69.64& 129.32& HET\\
17 24 13.32$+$27 56 55.2& 17.47& 0.78& 0.30& 0.11& 0.00& 0.26& 47& $-$60& 6& DA& 6131& 13.97& 14.37& 45.61& 16.48& 2.7m\\
17 28 07.29$+$26 46 20.1& 18.02& 0.97& 0.42& 0.16& 0.04& 0.23& $-$45& $-$255& 6& DA& 5619& 14.36& 14.85& 46.57& 57.15& 2.7m\\
20 41 28.99$-$05 20 27.7& 19.09& 1.65& 0.70& 0.26& 0.06& 0.26& $-$149& $-$29& 6& DC& 4673& 15.17& 15.98& 45.68& 32.87& MMT\\
20 42 59.23$+$00 31 56.6& 19.67& 1.65& 0.81& 0.30& 0.07& 0.37& $-$71& $-$244& 6& DC& 4201& 15.63& 16.57& 47.29& 56.96& HET\\
20 45 06.97$+$00 37 34.4& 19.43& 0.59& 0.25& 0.12& 0.00& 0.45& 32& $-$32& 6& DA& 6093& 14.00& 14.40& 117.73& 25.25& MMT\\
20 45 57.53$-$07 10 03.5& 19.08& 1.61& 0.68& 0.21& 0.12& 0.39& $-$73& $-$134& 6& DC& 4682& 15.16& 15.97& 47.90& 34.65& MMT\\
21 03 30.85$-$00 24 46.4& 18.22& 0.66& 0.27& 0.07& 0.02& 0.34& 61& $-$139& 6& DC& 6223& 13.91& 14.29& 68.61& 49.36& MMT\\
21 16 40.30$-$07 24 52.7& 17.93& 1.67& 0.69& 0.25& 0.05& 0.70& 111& $-$223& 6& DC& 4359& 15.47& 16.38& 25.89& 30.56& 2.7m\\
21 18 05.21$-$07 37 29.1& 19.85& 2.42& 0.98& 0.33& 0.11& 1.17& 115& $-$144& 5& DC& 3401& 16.55& 17.37& 43.44& 37.95& HET\\
21 18 58.65$+$11 20 17.7& 18.13& 0.77& 0.27& 0.09& 0.01& 0.42& 359& $-$13& 5& DA& 6086& 14.00& 14.40& 64.24& 109.38& 2.7m\\
21 25 01.48$-$07 34 56.0& 19.48& 0.74& 0.26& 0.08& 0.01& 0.56& 64& 13& 5& DA& 6063& 14.02& 14.42& 122.54& 37.93& MMT\\
21 36 43.08$-$07 06 38.2& 19.43& 0.60& 0.25& 0.08& $-$0.03& 0.20& 69& $-$11& 5& DA& 6520& 13.70& 14.05& 125.93& 41.71& MMT\\
21 47 52.10$-$08 24 36.8& 17.59& 0.75& 0.32& 0.14& $-$0.01& 0.24& 30& 153& 6& DA& 6010& 14.06& 14.47& 45.77& 33.82& 2.7m\\
21 54 30.69$+$13 00 26.7& 18.75& 1.55& 0.61& 0.25& 0.10& 0.41& 367& $-$73& 6& DZA& 4768& 15.08& 15.86& 43.76& 77.62& MMT\\
21 55 01.53$+$12 01 16.4& 18.47& 0.65& 0.25& 0.07& $-$0.06& 0.54& $-$76& $-$25& 4& DA& 6121& 13.98& 14.37& 79.14& 30.01& MMT\\
22 04 14.16$-$01 09 31.2& 19.88& 1.88& 0.83& 0.24& 0.11& 0.44& 112& $-$303& 6& DC& 4189& 15.65& 16.58& 52.87& 80.95& HET\\
22 22 33.90$+$12 21 43.0& 19.13& 1.95& 1.02& 0.36& 0.18& 0.41& 731& 198& 4& DC& 3448& 16.49& 17.33& 25.78& 92.53& HET\\
22 41 57.63$+$13 32 38.8& 17.36& 0.82& 0.35& 0.10& 0.02& 0.26& 61& $-$395& 6& DA& 5986& 14.07& 14.49& 40.90& 77.49& 2.7m\\
22 42 06.19$+$00 48 22.8& 19.38& 2.43& 0.91& 0.34& 0.08& 0.36& 132& $-$76& 6& DC& 3407& 16.55& 17.37& 29.12& 21.02& HET\\
22 54 08.64$+$13 23 57.2& 19.33& 2.04& 0.97& 0.33& 0.10& 0.26& 329& $-$199& 6& DC& 3356& 16.61& 17.41& 26.72& 48.71& HET\\
23 12 06.08$+$13 10 57.6& 17.45& 1.38& 0.56& 0.16& 0.08& 0.38& $-$132& $-$256& 6& DA& 5078& 14.80& 15.46& 28.40& 38.78& 2.7m\\
23 25 19.89$+$14 03 39.7& 16.30& 1.55& 0.57& 0.27& 0.09& 0.23& 336& 115& 6& DC& 4941& 14.92& 15.63& 14.80& 24.92& 2.7m\\
23 30 40.47$+$01 00 47.4& 17.36& 0.64& 0.17& 0.06& 0.02& 0.20& $-$255& $-$125& 6& DA& 6768& 13.54& 13.88& 53.63& 72.19& 2.7m\\
23 30 55.20$+$00 28 52.3& 19.77& 1.96& 0.89& 0.30& 0.11& 0.17& 151& 91& 6& DC& 4126& 15.71& 16.65& 44.42& 37.12& HET\\
23 37 07.68$+$00 32 42.3& 18.13& 1.01& 0.45& 0.14& $-$0.02& 0.18& 305& 162& 6& DA& 5629& 14.35& 14.84& 48.35& 79.15& 2.7m\\
23 40 41.47$-$11 06 36.9& 18.46& 0.53& 0.25& 0.74& 0.85& 0.15& 19& $-$87& 6& DA+M& 6612& 13.64& 13.99& 82.58& 34.86& MMT\\
23 42 45.75$-$10 01 21.4& 18.83& 1.52& 0.71& 0.26& 0.04& 0.16& $-$28& $-$95& 6& DA& 4719& 15.12& 15.92& 40.18& 18.86& MMT\\
23 50 42.52$-$08 46 18.9& 19.03& 1.04& 0.50& 0.22& 0.08& 0.18& 209& $-$139& 6& DA& 5298& 14.61& 15.20& 62.18& 73.97& HET\\
23 54 16.59$+$00 30 01.2& 19.25& 0.71& 0.20& 0.10& 0.04& 0.20& 53& 18& 5& DA& 6568& 13.67& 14.02& 119.10& 31.60& MMT
\enddata
\end{deluxetable}

\clearpage
\begin{deluxetable}{lcrrrrrrrrrr}
\tabletypesize{\tiny}
\tablecolumns{12}
\tablewidth{0pt}
\tablecaption{Spectroscopically Identified Subdwarf Stars}
\tablehead{
\colhead{Name (SDSS J)}&
\colhead{$g$}&
\colhead{$u-g$}&
\colhead{$g-r$}&
\colhead{$r-i$}&
\colhead{$i-z$}&
\colhead{$A_{\rm u}$}&
\colhead{$\mu_{\rm ra}$}&
\colhead{$\mu_{\rm dec}$}&
\colhead{Ep}&
\colhead{Dist22}&
\colhead{Source}
}
\startdata 
00 18 13.74$-$08 54 58.6& 19.48& 0.73& 0.24& 0.07& $-$0.03& 0.23& $-$41& $-$33& 5& 15.1& MMT\\
00 29 58.84$+$15 18 41.1& 19.20& 0.84& 0.29& 0.08& 0.07& 0.34& $-$46& $-$31& 4& 30.4& MMT\\
00 53 31.26$+$00 05 09.8& 19.38& 2.18& 1.09& 0.53& 0.23& 0.13& 137& $-$22& 6& 24.4& MMT\\
01 05 02.05$+$14 01 54.4& 19.40& 1.00& 0.43& 0.15& 0.08& 0.37& $-$11& 48& 5& 25.4& MMT\\
01 31 19.61$+$00 02 57.6& 17.93& 2.18& 1.14& 0.45& 0.25& 0.17& 230& $-$115& 6& 31.0& 2.7m\\
01 48 33.19$-$01 10 43.3& 17.98& 1.91& 0.85& 0.36& 0.15& 0.17& 152& $-$106& 6& 13.8& 2.7m\\
01 56 32.67$+$14 47 29.7& 17.48& 1.52& 0.66& 0.27& 0.07& 0.29& 111& $-$95& 5& 7.8& 2.7m\\
02 12 13.80$+$00 00 42.1& 18.81& 1.33& 0.51& 0.21& 0.10& 0.16& 81& 61& 5& 12.9& MMT\\
02 29 47.61$-$08 50 20.2& 16.80& 1.62& 0.74& 0.24& 0.14& 0.17& 151& $-$95& 6& 6.4& 2.7m\\
02 38 07.99$-$09 30 33.6& 18.61& 1.87& 0.77& 0.30& 0.19& 0.15& 80& $-$46& 6& 29.4& MMT\\
03 00 48.83$-$00 44 08.0& 17.21& 1.20& 0.60& 0.26& 0.19& 0.60& 83& $-$113& 6& 12.6& 2.7m\\
03 07 22.43$+$00 34 05.2& 15.68& 1.18& 0.49& 0.18& 0.08& 0.55& 156& $-$109& 6& 41.2& 2.7m\\
03 10 43.74$-$08 18 48.7& 17.49& 0.86& 0.25& 0.08& 0.03& 0.36& $-$32& 48& 5& 13.4& 2.7m\\
03 18 45.08$-$06 12 36.3& 16.56& 1.51& 0.73& 0.28& 0.18& 0.33& 159& $-$104& 6& 35.0& 2.7m\\
03 51 47.74$-$05 33 02.9& 15.94& 1.54& 0.74& 0.28& 0.12& 0.56& 237& $-$264& 6& 31.9& 2.7m\\
07 36 36.05$+$29 02 22.7& 17.41& 1.11& 0.34& 0.13& 0.05& 0.24& $-$18& $-$71& 5& 4.6& 2.7m\\
07 38 56.39$+$32 25 18.9& 17.52& 1.73& 0.80& 0.33& 0.16& 0.22& 5& $-$144& 6& 17.0& 2.7m\\
07 52 17.25$+$25 21 55.6& 16.79& 1.03& 0.27& 0.07& 0.00& 0.40& 146& 214& 4& 26.2& 2.7m\\
08 00 05.13$+$46 08 01.1& 16.64& 1.13& 0.55& 0.16& 0.11& 0.38& $-$2& 309& 4& 19.6& 2.7m\\
08 13 06.76$+$02 34 25.9& 17.15& 1.30& 0.45& 0.19& 0.09& 0.14& 309& $-$45& 5& 7.7& 2.7m\\
08 30 49.85$+$02 50 18.6& 18.24& 1.55& 0.53& 0.12& 0.12& 0.17& 45& 87& 6& 5.5& 2.7m\\
08 37 43.43$+$02 01 01.7& 17.91& 2.14& 0.78& 0.30& 0.11& 0.25& $-$15& $-$443& 4& 5.5& MMT\\
09 05 13.97$+$47 37 28.5& 18.67& 2.79& 1.33& 0.51& 0.31& 0.08& $-$159& $-$344& 6& 35.1& MMT\\
09 19 09.52$+$56 41 00.1& 17.03& 2.09& 0.70& 0.24& 0.14& 0.17& 1& $-$263& 5& 10.0& 2.7m\\
09 35 59.33$+$60 13 23.0& 17.22& 1.40& 0.55& 0.25& 0.09& 0.15& $-$4& $-$275& 4& 11.3& 2.7m\\
09 59 26.92$-$00 08 49.7& 16.82& 2.03& 0.79& 0.31& 0.09& 0.16& $-$315& 90& 5& 6.0& 2.7m\\
10 05 37.72$+$52 59 13.2& 17.30& 0.95& 0.24& 0.08& $-$0.01& 0.04& $-$3& 131& 5& 12.1& 2.7m\\
10 06 33.59$-$00 27 32.4& 17.95& 1.01& 0.47& 0.16& 0.08& 0.19& $-$237& $-$72& 5& 4.1& 2.7m\\
10 17 39.82$+$02 09 33.8& 16.79& 1.40& 0.59& 0.21& 0.12& 0.25& $-$240& 190& 4& 16.5& 2.7m\\
10 41 49.72$+$62 44 55.6& 16.69& 0.98& 0.29& 0.08& 0.05& 0.03& 55& 161& 5& 12.2& 2.7m\\
10 51 57.48$+$02 03 00.4& 19.74& 1.97& 1.05& 0.46& 0.28& 0.22& 166& $-$105& 5& 32.0& MMT\\
11 07 23.94$+$62 26 06.0& 16.29& 0.97& 0.33& 0.11& 0.03& 0.05& 46& 239& 4& 25.4& 2.7m\\
11 13 27.37$+$58 58 48.5& 17.09& 2.32& 1.24& 0.48& 0.31& 0.05& $-$296& $-$472& 6& 59.0& 2.7m\\
11 22 04.68$+$65 53 59.7& 17.54& 1.08& 0.36& 0.17& 0.04& 0.06& $-$63& $-$152& 5& 39.5& 2.7m\\
11 31 02.89$+$66 57 51.2& 16.66& 1.64& 0.52& 0.25& 0.11& 0.05& $-$116& 206& 5& 42.9& 2.7m\\
11 35 12.08$+$03 28 41.2& 19.24& 2.80& 1.18& 0.47& 0.31& 0.11& $-$163& $-$255& 6& 20.1& MMT\\
11 52 06.86$+$67 02 04.2& 16.33& 0.87& 0.21& 0.08& 0.02& 0.06& $-$117& 143& 4& 19.8& 2.7m\\
12 09 45.96$+$63 02 43.9& 16.96& 0.86& 0.25& 0.08& 0.01& 0.10& $-$71& 196& 5& 26.5& 2.7m\\
12 50 47.35$+$03 26 52.8& 18.28& 2.17& 0.98& 0.43& 0.19& 0.16& $-$97& $-$221& 4& 12.5& 2.7m\\
12 52 26.29$+$02 28 38.8& 16.78& 1.44& 0.49& 0.16& 0.05& 0.16& 155& $-$100& 5& 19.1& 2.7m\\
13 05 07.26$+$04 14 08.1& 16.81& 0.92& 0.32& 0.12& 0.02& 0.13& $-$135& $-$47& 5& 9.3& 2.7m\\
13 42 33.44$+$58 00 19.7& 17.16& 1.08& 0.32& 0.13& 0.04& 0.04& 15& 196& 5& 15.7& 2.7m\\
14 06 35.95$+$61 53 35.6& 19.70& 1.96& 1.03& 0.42& 0.25& 0.07& $-$84& 296& 5& 36.2& HET\\
16 37 16.86$+$45 17 01.8& 17.79& 0.95& 0.33& 0.08& 0.05& 0.06& $-$83& $-$414& 4& 13.6& 2.7m\\
17 02 06.35$+$31 47 49.9& 17.71& 1.57& 0.58& 0.19& 0.13& 0.21& 246& $-$99& 5& 9.6& 2.7m\\
17 17 37.03$+$62 34 48.1& 18.26& 1.10& 0.32& 0.16& 0.00& 0.12& $-$36& 43& 6& 28.5& 2.7m\\
17 19 18.54$+$29 15 38.6& 17.52& 1.55& 0.59& 0.14& 0.08& 0.20& 72& 142& 5& 11.1& 2.7m\\
17 35 52.77$+$57 38 14.3& 17.88& 2.07& 1.00& 0.40& 0.19& 0.33& $-$89& 156& 6& 20.0& 2.7m\\
20 58 59.24$-$05 57 03.8& 19.16& 1.51& 0.59& 0.19& 0.08& 0.23& $-$51& 51& 4& 12.1& MMT\\
21 34 27.61$-$08 15 11.0& 18.41& 1.34& 0.49& 0.15& 0.08& 0.18& $-$75& 19& 5& 23.0& MMT\\
21 42 39.11$-$00 55 50.3& 18.69& 2.41& 1.19& 0.47& 0.26& 0.27& 103& $-$173& 6& 13.6& MMT\\
21 48 19.02$+$00 39 43.2& 18.30& 0.99& 0.35& 0.10& 0.01& 0.81& $-$55& 79& 6& 9.7& MMT\\
22 14 28.87$+$13 53 41.9& 17.66& 1.04& 0.39& 0.14& 0.06& 0.33& 90& 50& 6& 8.7& 2.7m\\
22 54 49.68$+$12 59 22.6& 19.23& 1.75& 1.00& 0.31& 0.21& 0.20& 226& $-$125& 5& 7.2& HET\\
22 58 03.60$-$10 07 02.0& 19.33& 2.73& 1.08& 0.57& 0.29& 0.20& 108& $-$145& 6& 28.8& MMT
\enddata
\tablecomments{Proper motions for stars that have not been detected in 1 or 2 USNO-B epochs (Ep$<$6) or
that have a neighbor brighter than 22 mag within 7 arcsec (Dist22$<7$) are unreliable.}
\end{deluxetable}

\clearpage
\begin{figure}
\plotone{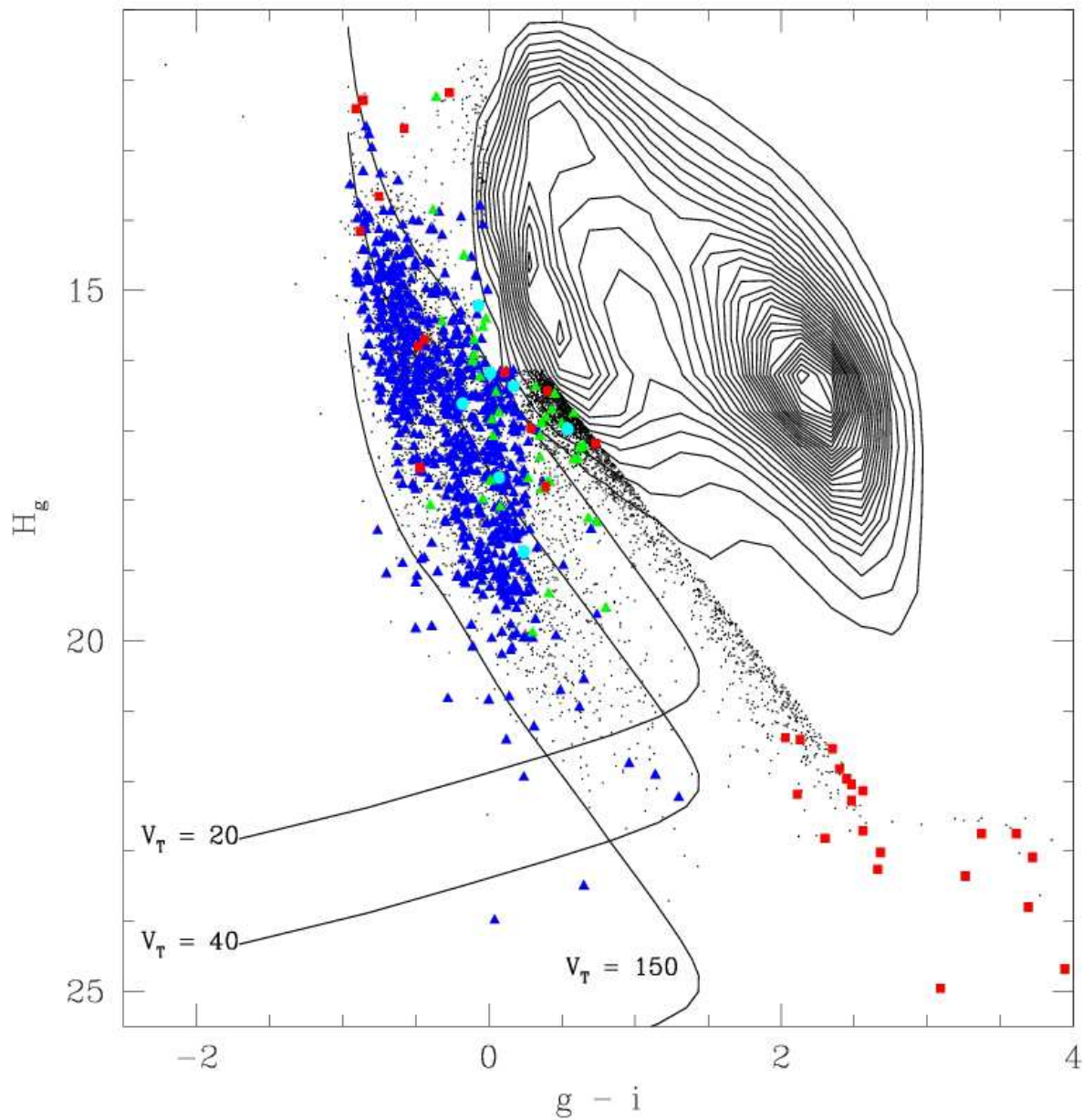}
\caption{The reduced proper motion diagram for stars in the SDSS DR2.
Individual stars are plotted only in the region of interest for white dwarfs, the remaining
stars are represented by the contours. Previously known white dwarfs, white dwarf plus late type star
binaries, subdwarfs, and quasars are shown as blue triangles, green triangles, red squares, and cyan circles,
respectively.
White dwarf cooling curves for different tangential velocities are shown as solid
lines. The $V_{\rm T}=$ 20--40 km s$^{-1}$ curves mark the expected location of disk white dwarfs,
whereas the $V_{\rm T}=$ 150 km s$^{-1}$ curve represents the halo white dwarfs.}
\end{figure}

\clearpage
\begin{figure}
\plotone{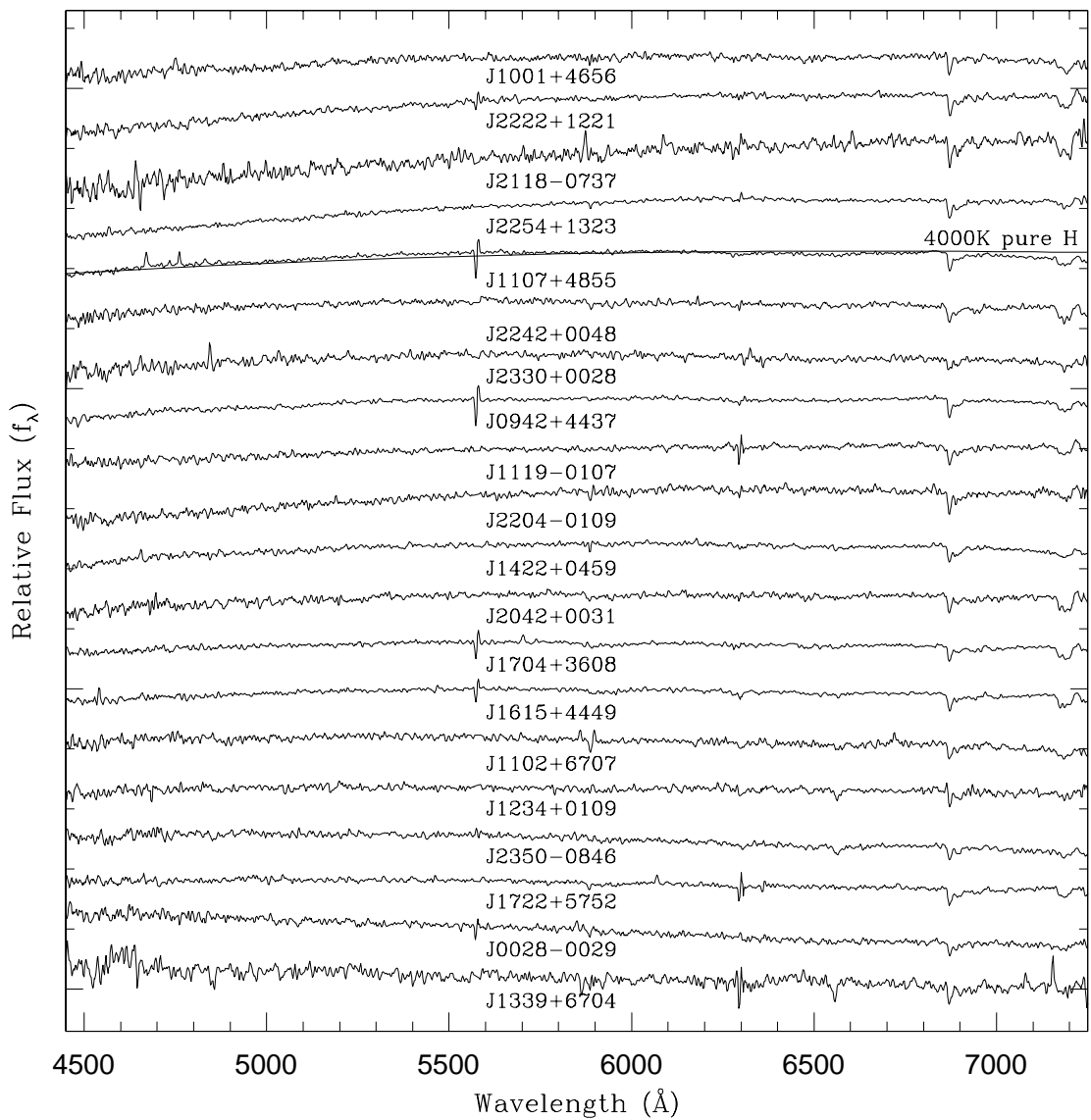}
\caption{Optical spectra for white dwarfs observed at the Hobby--Eberly Telescope. The spectra
are normalized at 5500 \AA, and are shifted vertically from each other by arbitrary
units. Synthetic spectrum of a 4000 K pure H atmosphere white dwarf (D. Saumon,
private communication) is also shown. The $g-r$ color increases from bottom to top.}
\end{figure}

\clearpage
\begin{figure}
\plotone{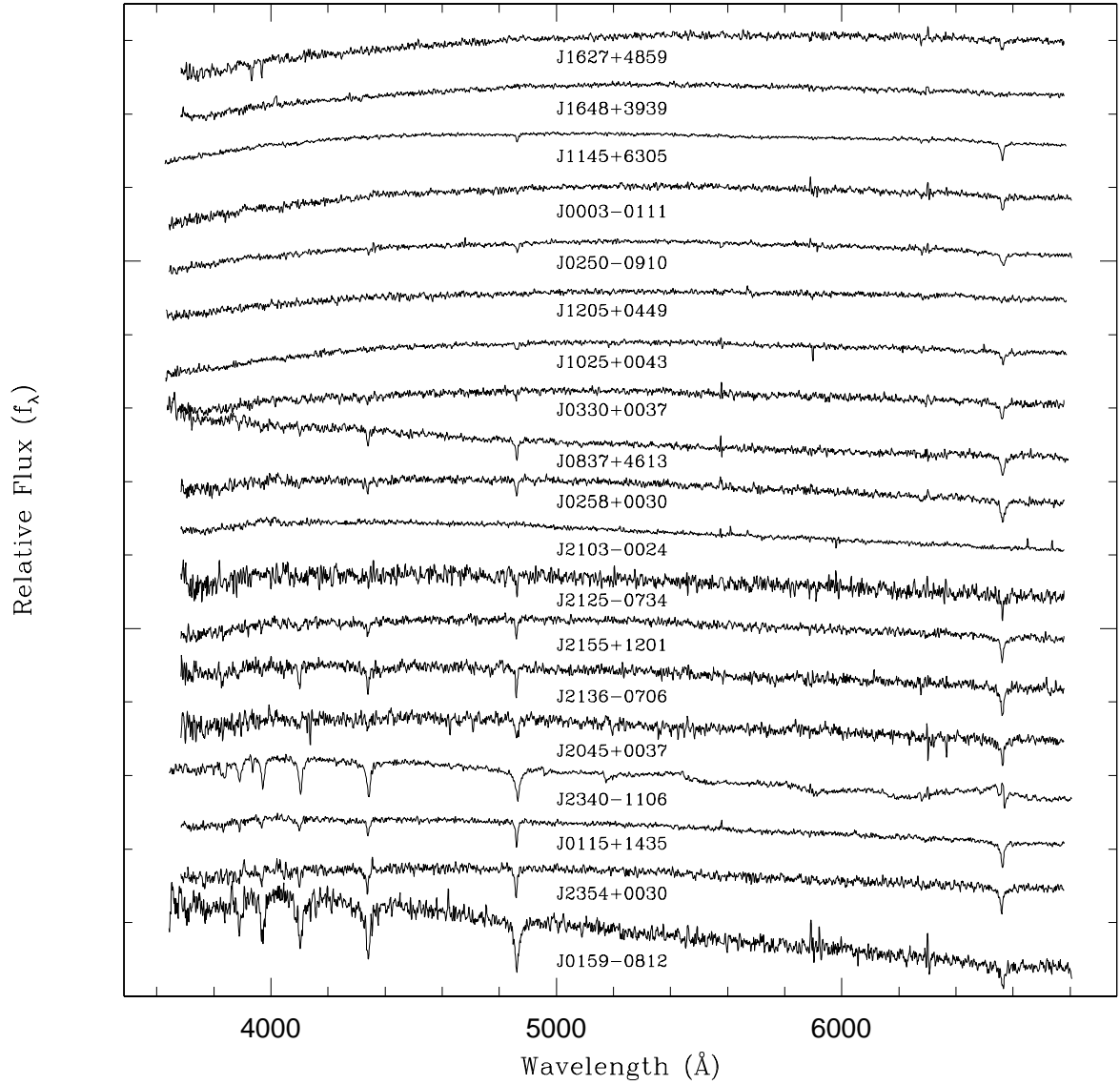}
\caption{Same as figure 2, but for white dwarfs observed at the MMT.}
\end{figure}

\clearpage \begin{figure}
\plotone{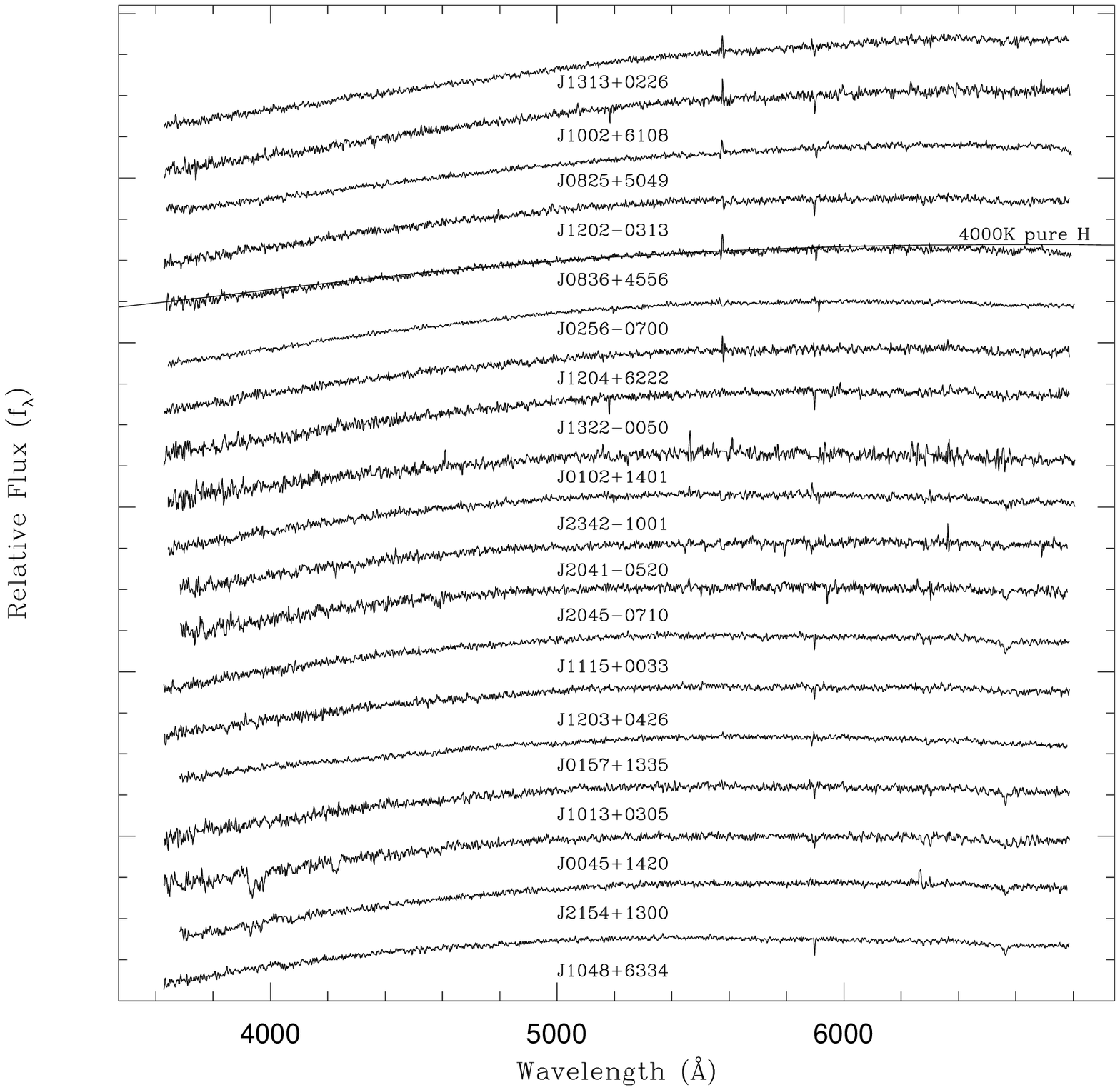}
\begin{flushright}
Figure 3b
\end{flushright}
\end{figure}

\clearpage
\begin{figure}
\plotone{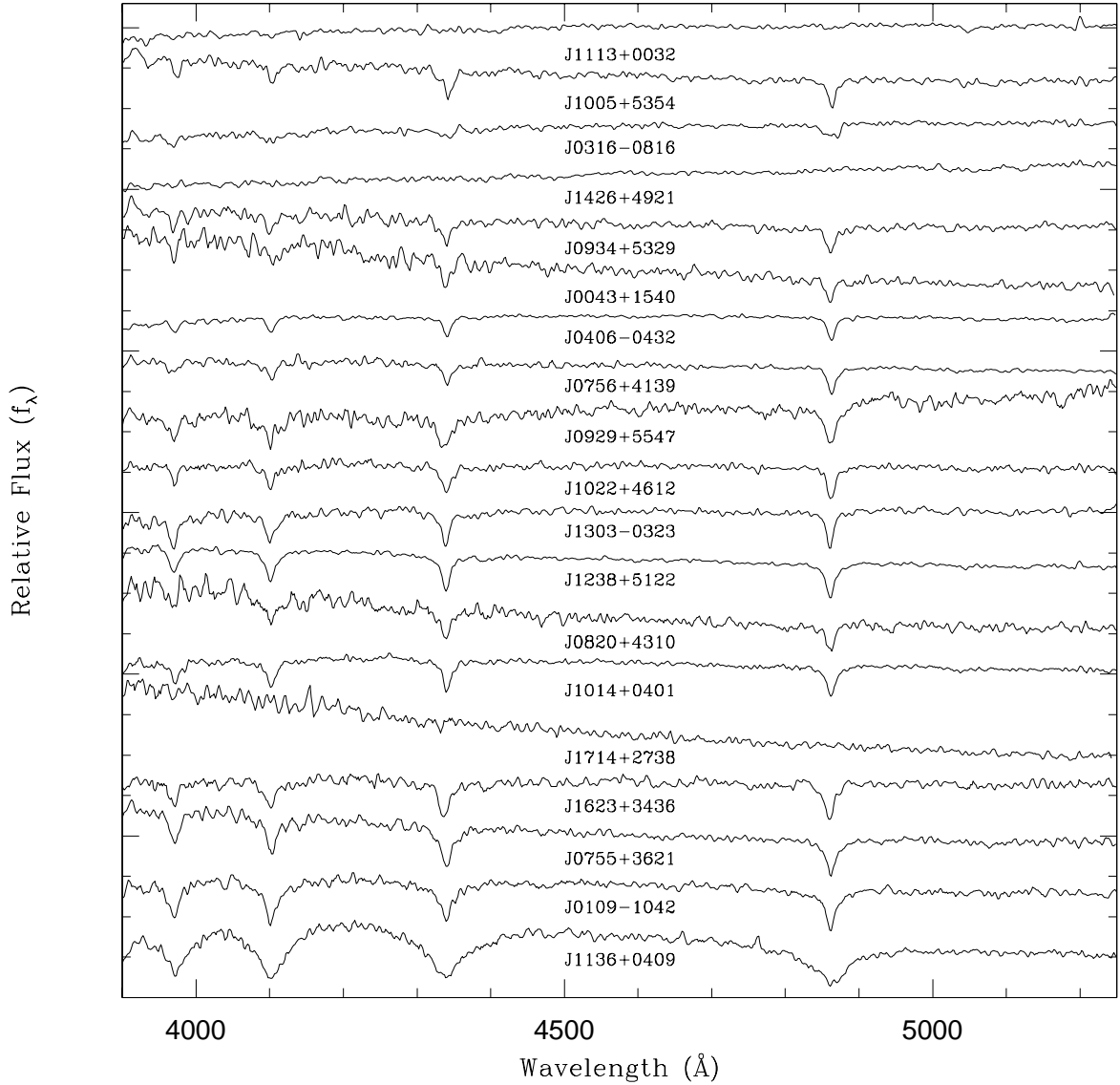}
\caption{Optical spectra for white dwarfs observed at the McDonald 2.7m Harlan Smith Telescope. The spectra
are normalized at 4600 \AA, and are shifted vertically from each other by arbitrary units.
The $u-r$ color increases from bottom to top.}
\end{figure}

\clearpage
\begin{figure}
\plotone{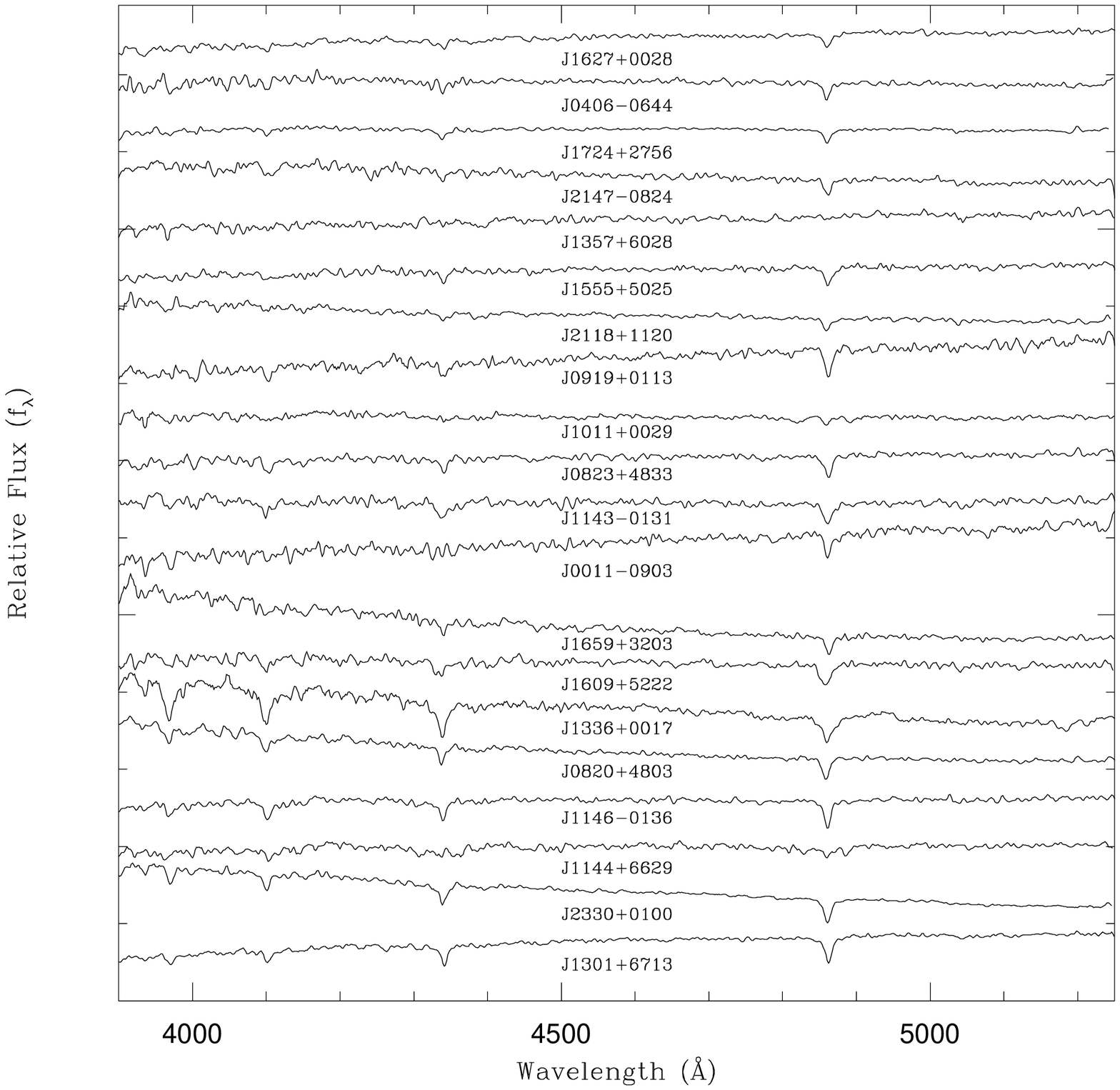}
\begin{flushright}
Figure 4b
\end{flushright}
\end{figure}

\clearpage
\begin{figure}
\plotone{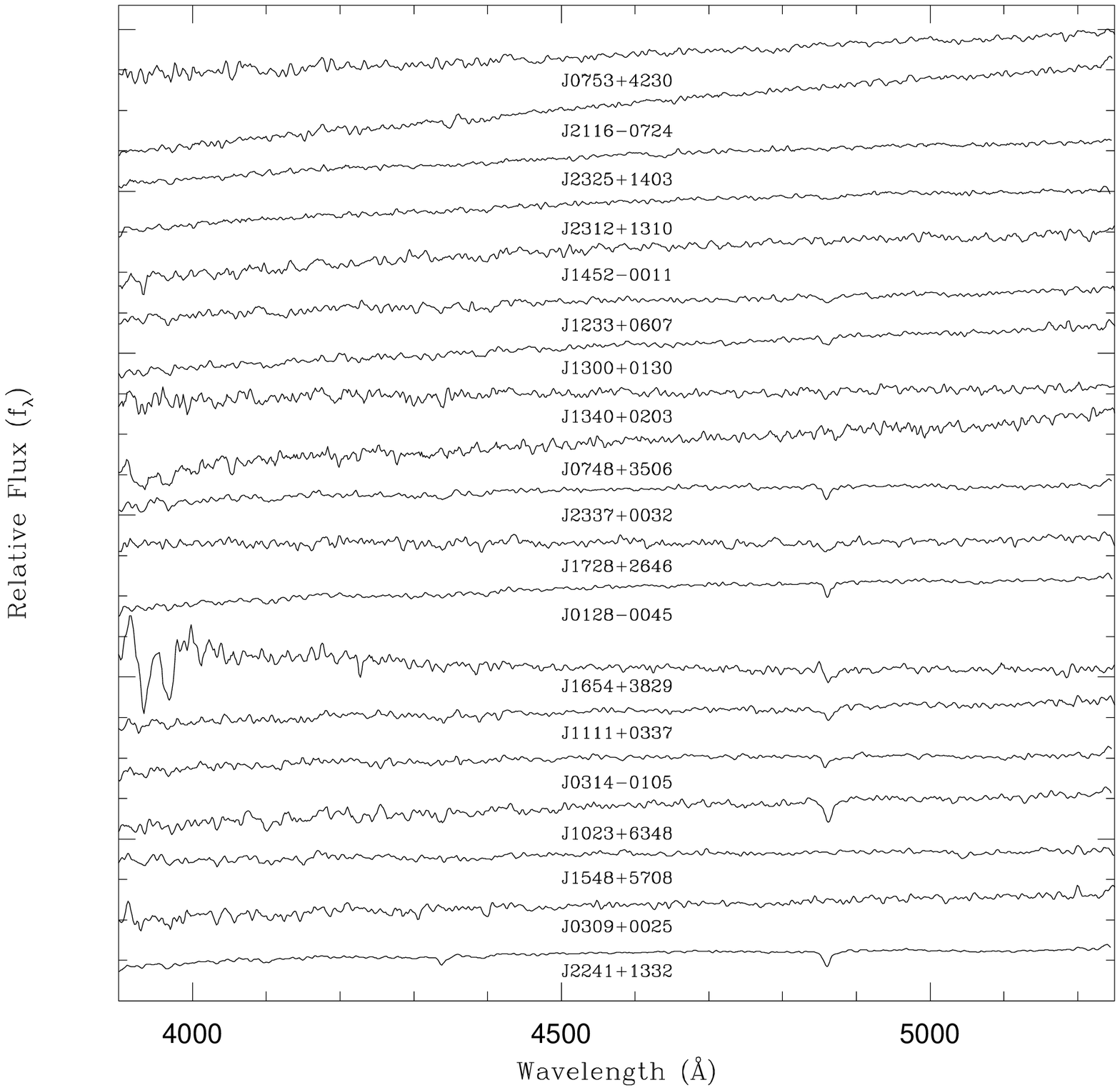}
\begin{flushright}
Figure 4c
\end{flushright}
\end{figure}

\clearpage
\begin{figure}
\plotone{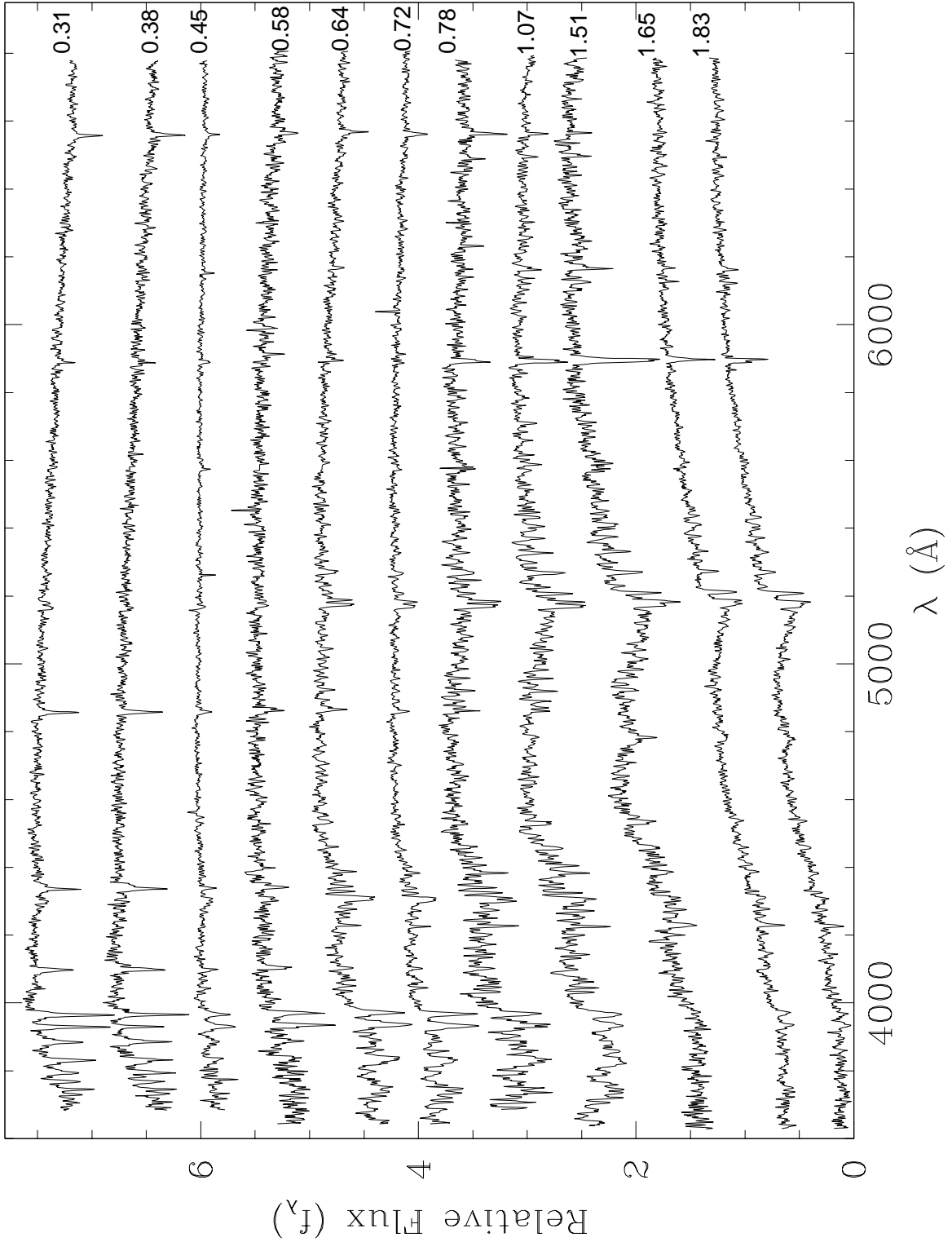}
\caption{Optical spectra of several subdwarf stars observed at the MMT. The spectra are ordered in increasing $g-i$ color, which
is given on the right edge of each spectrum.}
\end{figure}

\clearpage
\begin{figure}
\plotone{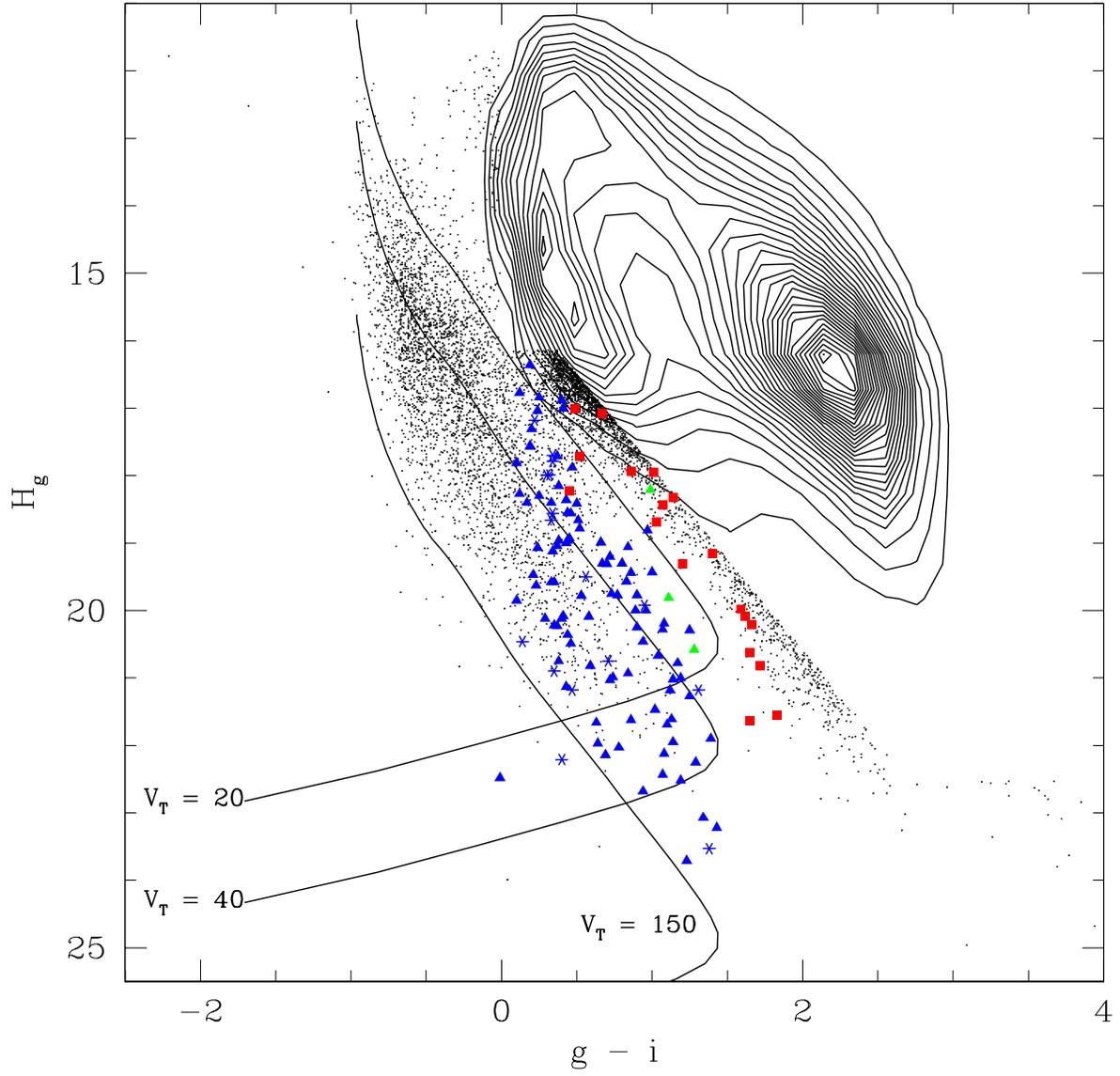}
\caption{Same as figure 1, but for the spectroscopically confirmed white dwarfs, white dwarf +
late type star binaries, subdwarfs, and quasars found in our study. White dwarfs that did not
meet our criteria for reliable proper motions are plotted as blue asterisks.}
\end{figure}

\clearpage
\begin{figure}
\plotone{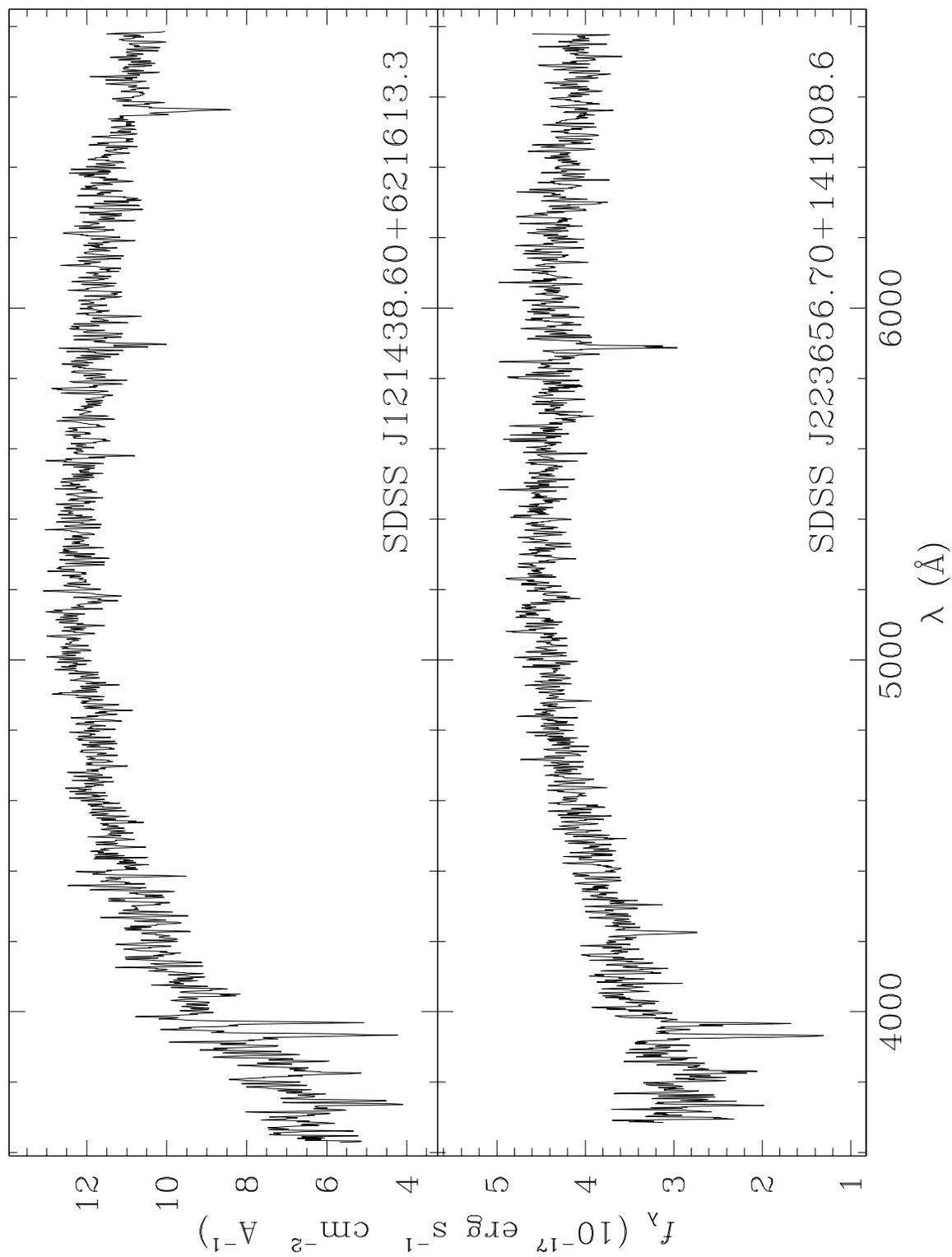}
\caption{MMT spectra for two uncertain white dwarfs / subdwarfs.}
\end{figure}

\clearpage
\begin{figure}
\plotone{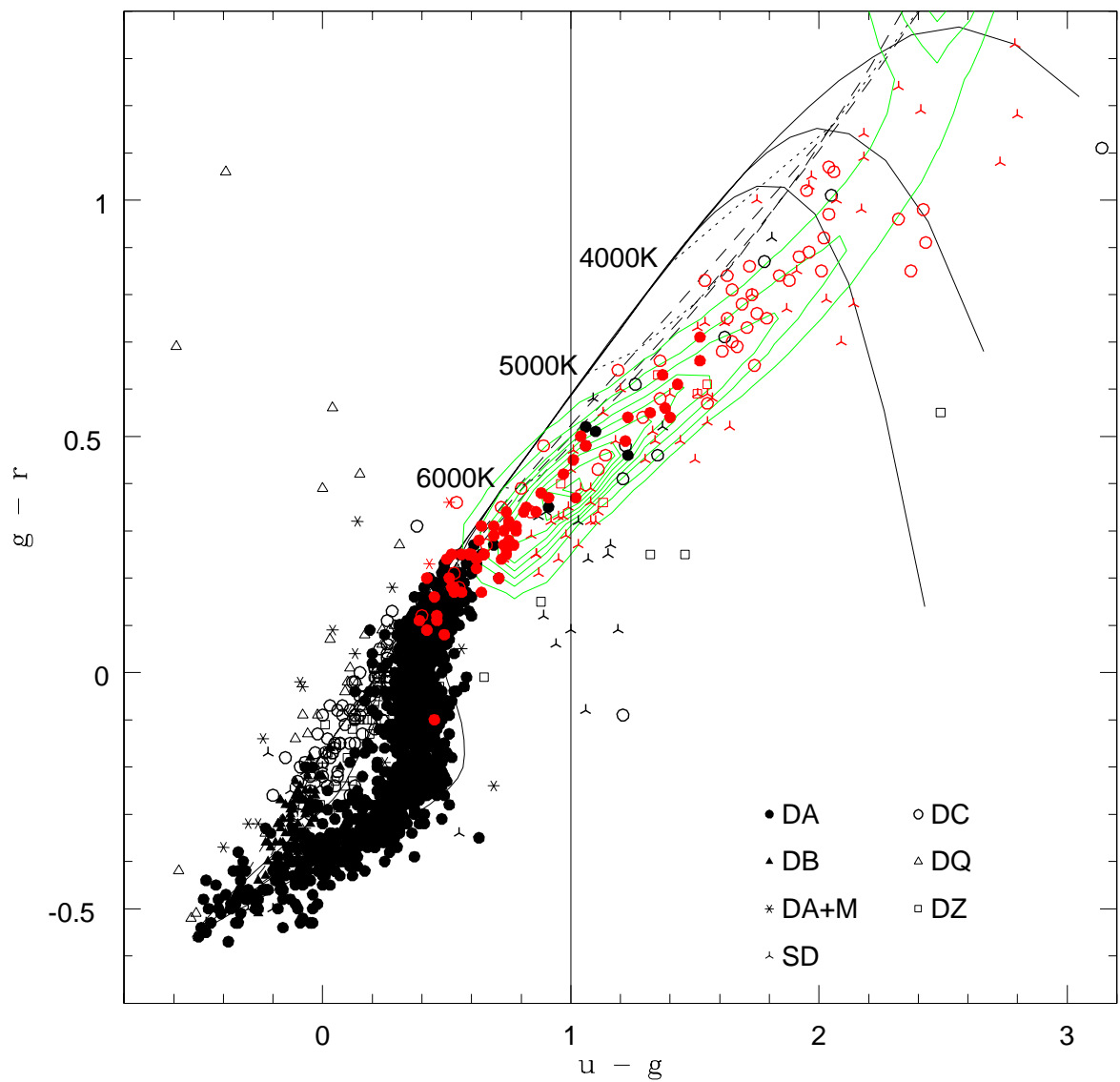}
\caption{Color-color diagrams showing the white dwarfs and subdwarfs from our study (red symbols) and
the literature (black symbols). Different
types of white dwarfs are shown with different symbols.
The contours represent objects without spectroscopic confirmation.
The curves show the colors
of white dwarf model atmospheres (P. Bergeron, private communication) of pure H (solid curves) and
pure He (dashed curves) with log $g=$ 7, 8, and 9, where the log $g=$ 9 curve is the
bottom and log $g=$ 7 is the top curve in panel $a$. The dotted lines with labels
connect models with the same effective temperature.}
\end{figure}

\clearpage
\begin{figure}
\plotone{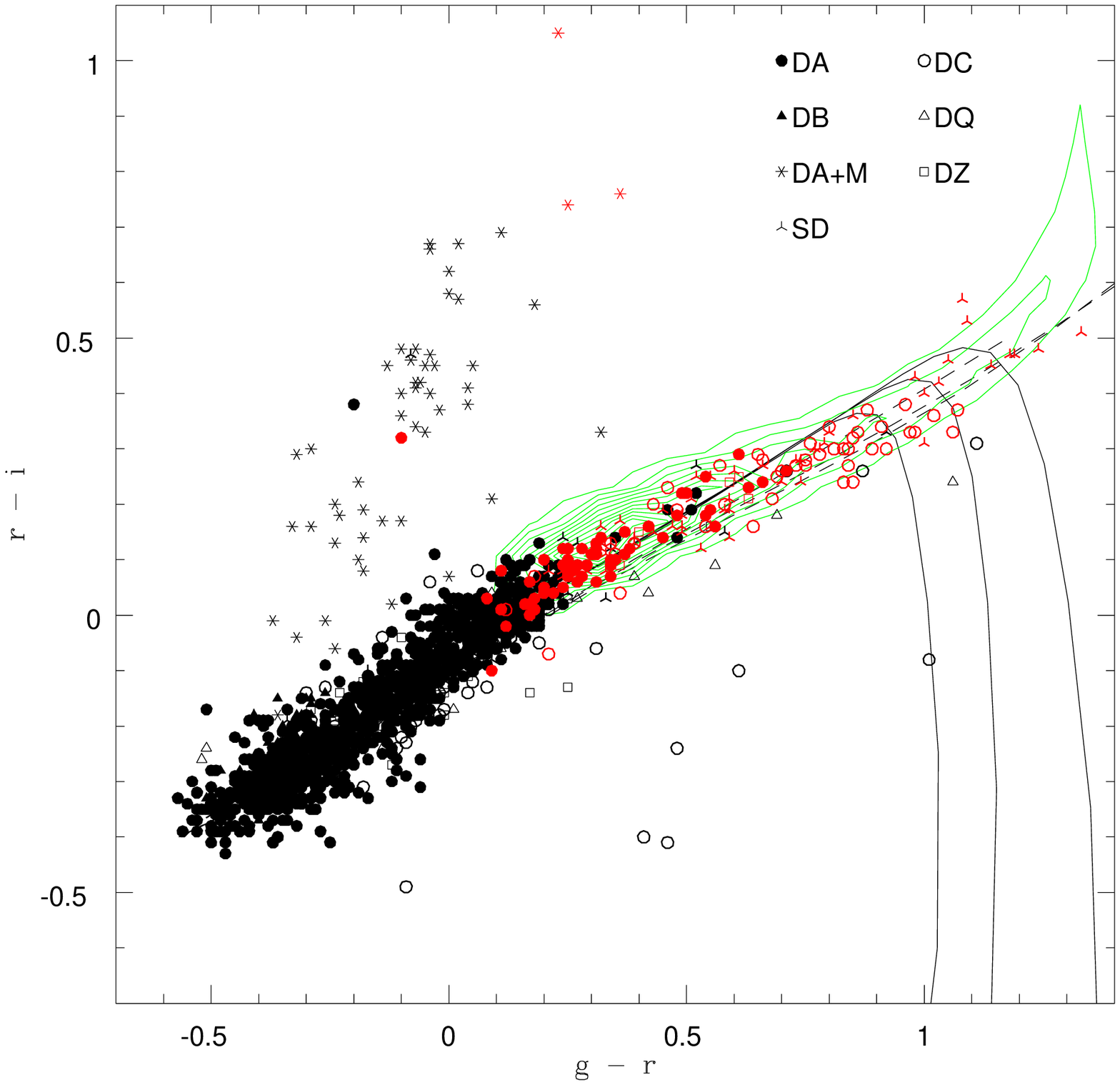}
\begin{flushright}
Figure 8b
\end{flushright}
\end{figure}

\clearpage
\begin{figure}
\plotone{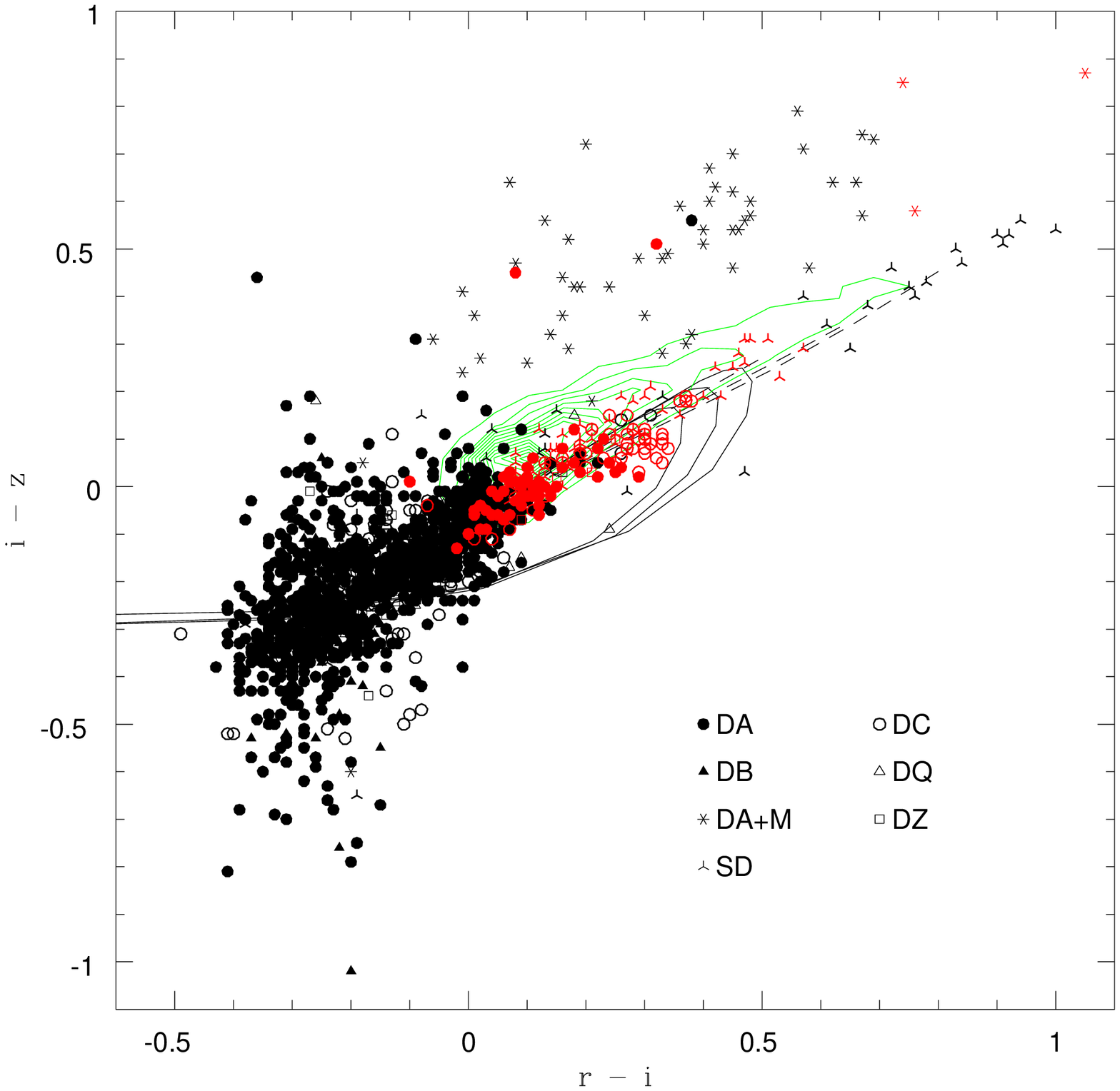}
\begin{flushright}
Figure 8c
\end{flushright}
\end{figure}

\clearpage
\begin{figure}
\plotone{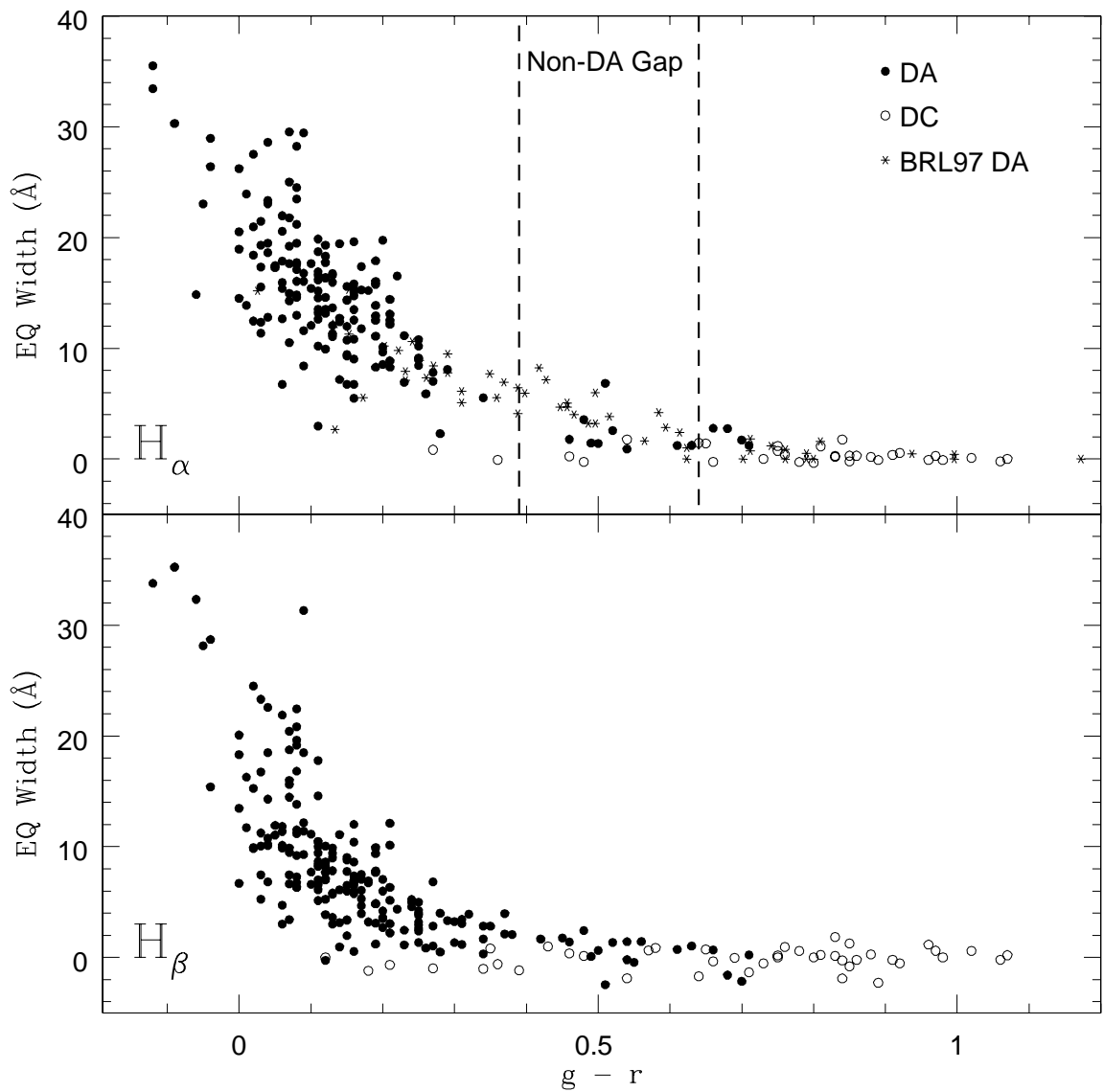}
\caption{Equivalent width measurements of H$\alpha$ and H$\beta$ as a function of $g-r$ for the DA and DC
white dwarfs in our sample. The top panel also includes H$\alpha$ equivalent width measurements
of the Bergeron, Ruiz \& Leggett (1997) DA sample. The predicted non-DA gap for log $g=8$ DA white
dwarfs is marked by the dashed lines.
Since McDonald 2.7m spectra do not cover H$\alpha$, we
do not have H$\alpha$ equivalent width measurements for some of our objects.}
\end{figure}

\clearpage
\begin{figure}
\plotone{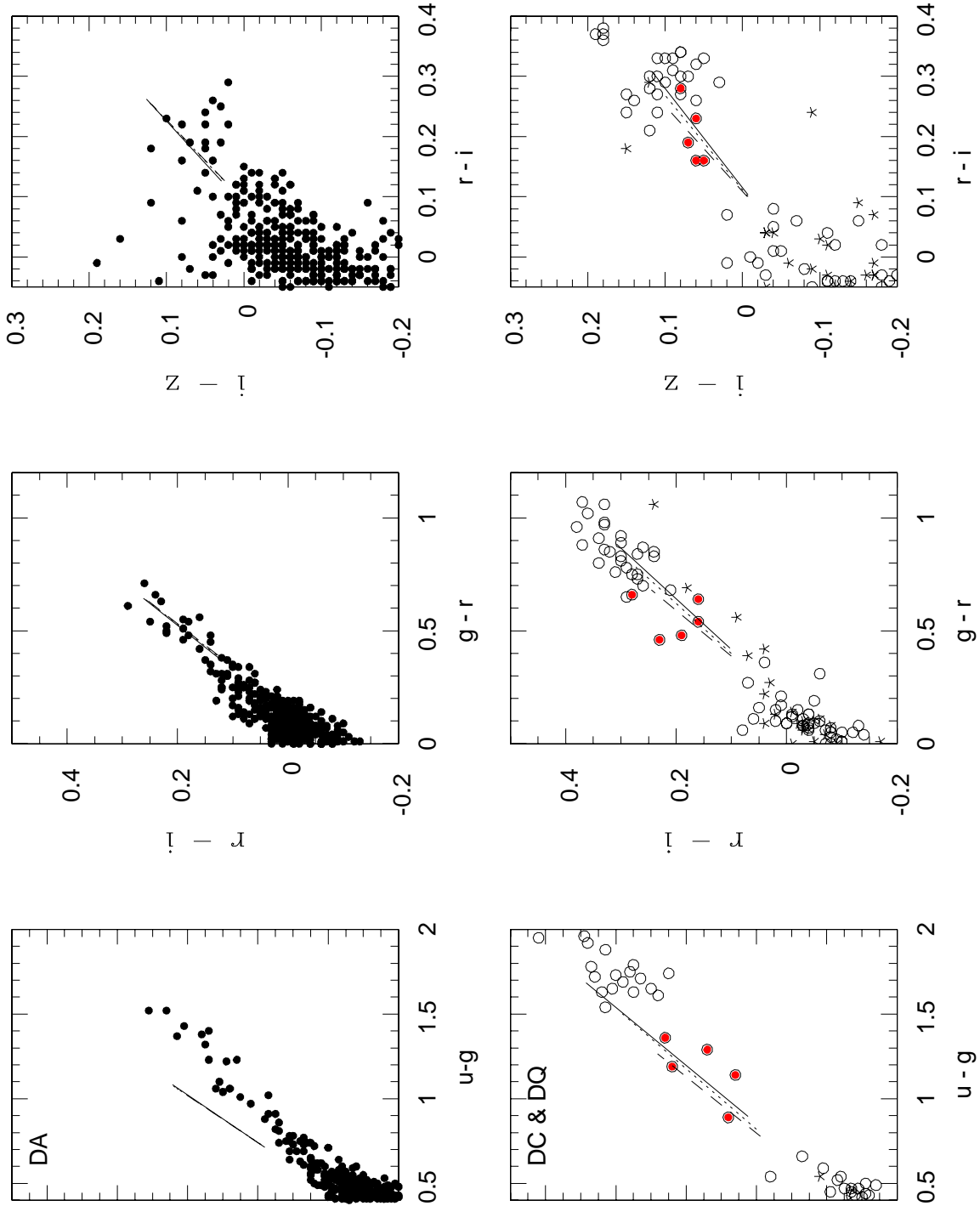}
\vspace{-1.2in}
\caption{The $u-g$ vs. $g-r$ (left panel), $g-r$ vs. $r-i$ (middle panel), and
$r-i$ vs. $i-z$ (right panel) color-color diagrams for white dwarfs in our sample.
Upper panels show DA white dwarfs, whereas lower panels show DC (open circles) and DQ (star symbols) white dwarfs.
The pure H (DA panels) and pure He (non-DA panels)
model sequences with $6000K\ga T_{\rm eff}\ga5000K$ are also shown for log $g=7$ (solid line), log $g=8$
(dashed line), and log $g=9$ (dotted line). Probable DC white dwarfs in the non-DA gap are shown as filled red circles.}
\end{figure}

\clearpage
\begin{figure}
\plotone{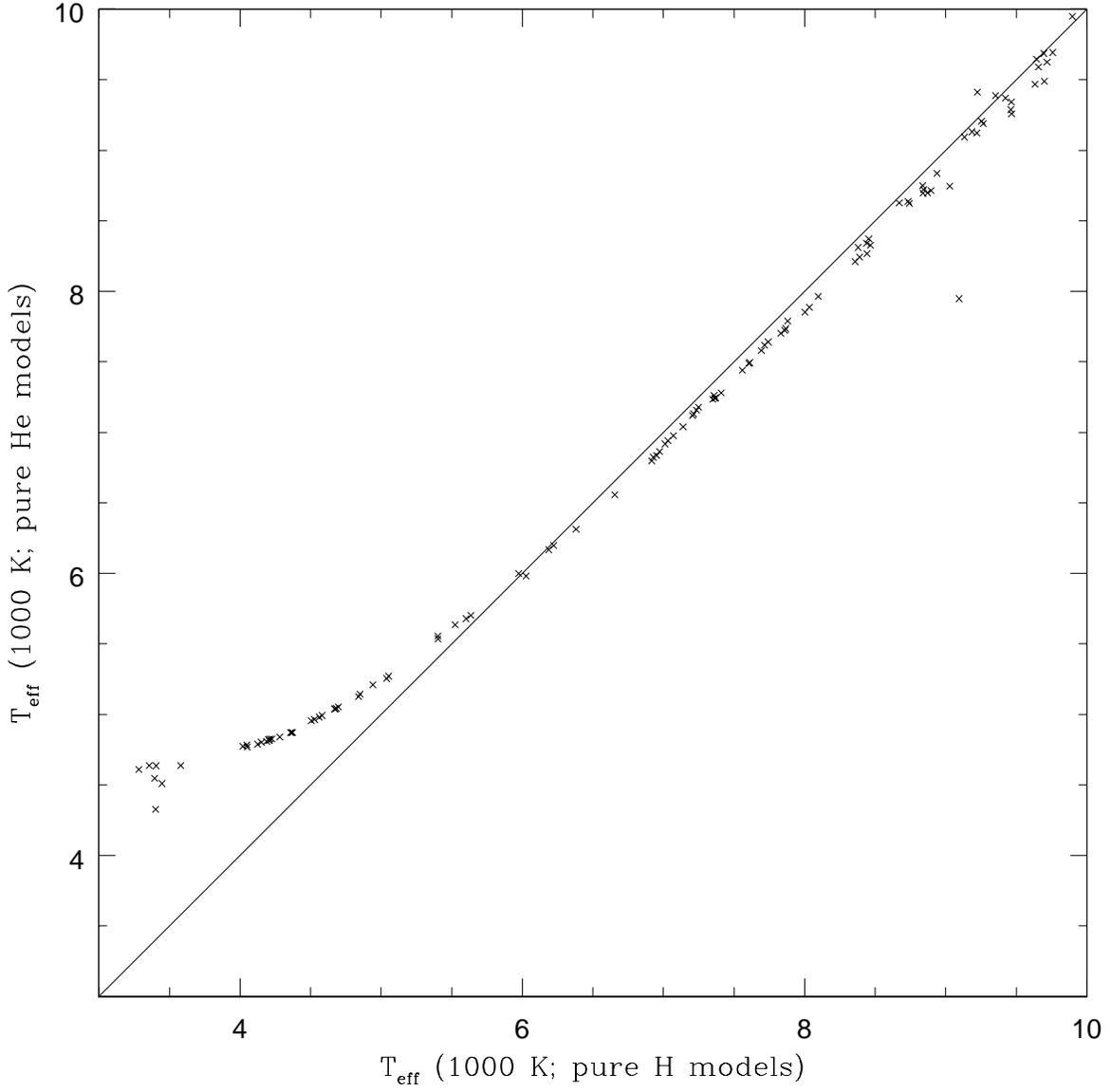}
\caption{Effective temperatures for our sample of DC white dwarfs using pure H or pure He white dwarf model atmospheres.
The effects of using a hydrogen-rich or a helium-rich composition to estimate temperatures become
significant below $\sim5500K$.}
\end{figure}

\end{document}